\documentclass[fleqn,11pt,letterpaper]{article}
\usepackage[utf8]{inputenc}
\usepackage{caption}
\usepackage{tikz} 
\usepackage{tikz}
\usepackage{setspace}
\usepackage{float}
\usetikzlibrary{decorations.pathreplacing, positioning}

\usetikzlibrary{decorations.pathreplacing, positioning}

\usepackage{graphicx}
\usepackage{subcaption}
\usepackage[section]{placeins}
\usepackage{amsmath}
\usepackage{amssymb}
\usepackage{booktabs}
\usepackage{geometry}
\usepackage{array}
\usepackage{algorithm}
\usepackage{algpseudocode}
\usepackage{longtable}
\usepackage{graphicx}
\usepackage{multirow}

\usepackage[table]{xcolor}

\usepackage{tikz}
\usepackage{caption}
 \def\spacingset#1{\renewcommand{\baselinestretch}{#1}\small\normalsize}  

\title{Uncovering Multimorbidity Patterns Through a Novel Randomized Clustering Algorithm}

\author{Ali Jafari and Hari Balasubramanian\\ University of Massachusetts Amherst}
\date{}

\usepackage[style=apa, backend=biber, apamaxprtauth=99]{biblatex}
\DeclareLanguageMapping{american}{american-apa}
\usepackage{csquotes}
\usetikzlibrary{positioning}

\begin{document}

\maketitle

\begin{abstract}

Patients with several chronic conditions rarely share the same combination: in four years
of the nationally representative Medical Expenditure Panel Survey (MEPS), we find 32,789
distinct combinations of two to seven conditions, and over 90\% of those involving four or
more occur in a single patient. Methods that search for common combinations discard most of
this heterogeneity. We present the Randomized Length Sequential Clustering Algorithm
(RLSCA), which links diseases by lift, a pairwise association measure counted only above a
user-set support threshold. A mixed-integer program repeatedly selects and removes the
group of conditions with the greatest internal lift, up to a random size; 100 iterations
yield a database of clusters scored by size, rediscovery frequency, and body systems
spanned. In MEPS 2019, stratified by age and sex, RLSCA recovers known groupings at a high
threshold, such as diabetes with lipid disorders and hypertension, and at a lower threshold
returns clusters of up to eight conditions spanning five body systems. Frequent itemset
mining, at its lowest feasible support, finds no set above five conditions; hierarchical
clustering cannot control cluster size. With a clinical large language model, the database
could suggest which further conditions a patient is at risk of.

\end{abstract}

\vspace{0.5em}

\spacingset{1.5} 
\renewcommand{\thesection}{\arabic{section}.}
\renewcommand{\thesubsection}{\thesection\arabic{subsection}.}
\renewcommand{\thesubsubsection}{\thesubsection\arabic{subsubsection}.}

\section{Introduction}

Multimorbidity refers to the coexistence of two or more diseases  (\cite{skou2022multimorbidity}) in an individual and has been recognized as a global challenge. Since many of the co-existing diseases tend to be chronic conditions, the term \textit{multiple chronic conditions (MCC)} is also used in this context. According to a 2017 RAND report, 40\% of U.S. adults have no chronic conditions and account for only 10\% of total healthcare costs, while 12\% have five or more chronic conditions and are responsible for 41\% of healthcare spending (\cite{buttorff2017multiple, balasubramanian2024interventions}). Thus a small percent of patients with multiple conditions have a disproportionate impact on costs and utilization, motivating the need to understand disease patterns in this subpopulation. 

One of the main challenges in finding patterns among patients with multiple conditions is the high degree of heterogeneity observed in condition combinations. To illustrate this heterogeneity, consider the chronic conditions of three example patients from the Medical Expenditure Panel Survey (MEPS) -- the dataset we use in this study -- reproduced from \textcite{balasubramanian2024interventions}. 

\begin{itemize}
    \item \textbf{Patient A (5 conditions)}: Lipid Metabolism Disorders; Anxiety; Depression; Non-Trau\-matic Joint Disorders; Spondylosis/Intervertebral Disc Disorders/Other Back Issues
    \item \textbf{Patient B (5 conditions)}: Diabetes; Lipid Metabolism Disorders; High Blood Pressure; Coronary Atherosclerosis/Heart Disease; Connective Tissue Disease
    \item \textbf{Patient C (6 conditions)}: Diabetes; Lipid Metabolism Disorders; High Blood Pressure; Chronic Obstructive Pulmonary Disease (COPD) and Bronchiectasis; Coronary Atherosclerosis/Heart Disease; Spondylosis/Intervertebral Disc Disorders/Other Back Issues
\end{itemize}

We observe that while the three patients do have common subsets of diseases -- for instance, Patients B and C share a subset of four conditions: Diabetes; Lipid Metabolism Disorders; High Blood Pressure; and Coronary Atherosclerosis/Heart Disease -- each patient's combination of conditions is unique. In a four-year extract of the MEPS dataset (2016-2019), we found 32,789 unique condition combinations involving 2-7 diseases. Over 90\% of combinations involving 4 or more conditions are unique to a single patient (i.e., they are not found in other patients). Yet, paradoxically, there are so many combinations  that they add up to a high combined prevalence in the population: 19\% of U.S. adults had four or more chronic conditions in 2014 (\cite{buttorff2017multiple, balasubramanian2024interventions}). \textcite{sorace2013understanding} aptly call this phenomenon \textit{“the high prevalence of low-prevalence chronic disease combinations”}. This pattern has been found in Medicare data as well (\cite{sorace2011complexity}). 

Multimorbidity also offers important clues on how diseases that are seemingly disparate and belong to different body and organ systems are related to each other. For instance, the Framingham Heart Study (\cite{kannel1979diabetes}) established that diabetes (endocrine and metabolic system), high blood pressure (circulatory system), and heart conditions are closely linked to each other. More recently, the health and diversity of the gut microbiome (digestive health) has been linked to mental health conditions (\cite{valles2019neuroactive}); and infection with the Epstein-Barr virus has been linked to multiple sclerosis, a chronic inflammatory demyelinating disease of the central nervous system (\cite{bjornevik_longitudinal_2022}). These discoveries highlight the need for a data-driven approach to discover as-yet-unknown relationships between diseases. Finding such patterns  helps with the screening, prevention and treatment options for patients with multiple conditions. It also identifies prevention strategies for the general population who do not yet have the diseases -- that is, it prevents the development of risk factors for disease in the first place. A prominent -- and widely adopted -- example is the emphasis on lifestyle changes (diet and physical activity) to prevent metabolic and cardiovascular conditions.   

The principal contribution of this paper is a new randomized integer-programming based clustering methodology that will help clinicians identify the most related disease clusters in these heterogeneous condition combinations.  Unlike off-the-shelf clustering methods such as hierarchical agglomerative clustering, exploratory factor analysis, latent class analysis, which have been widely applied in the multimorbidity context, our method is highly customized to extract disease-relatedness patterns.  Like all other methods, ours can be used to create a single set of final clusters. However, a single clustering solution is often inadequate, since diseases from different body systems can co-occur in numerous ways.  The randomized approach in our algorithm systematically explores and identifies a wide range of possible disease connections across different age and gender groups. We illustrate the results of our methodology using patient-level data in MEPS. Because the MEPS is nationally representative, our findings constitute perhaps the most detailed analysis of multiple disease patterns in the US population.

The rest of the paper is organized as follows:

\begin{itemize}
    \item \textbf{Section 2 -- Literature Review:} Reviews key studies on multimorbidity clustering, identifies gaps in existing methods, and explains contributions of this paper in relation to existing work.
    
    \item \textbf{Section 3 -- Methodology:} Describes the dataset, preprocessing steps, and the development of the Randomized Length Sequential Clustering Algorithm (RLSCA). This section concludes with our vision for integrating RLSCA into AI-driven clinical support systems.
    
    \item \textbf{Section 4 -- Alternative Clustering Approaches:} Adapts and describes baseline methodologies, including frequent itemset mining (Apriori/FP-growth) and hierarchical clustering, to provide a comparative framework for multimorbidity pattern discovery.
    
    \item \textbf{Section 5 -- RLSCA Results:} Presents the clustering outcomes across various age and gender subgroups, evaluates cluster characteristics (e.g., length, diversity, frequency), and validates the results of the RLSCA with what is known clinically.

    \item \textbf{Section 6 -- Conclusions and Future Research:} Summarizes the study's contributions and proposes extensions for future multimorbidity modeling.
\end{itemize}

\section{Literature Review}

Multimorbidity has been well studied in the clinical literature. The most common kind of analysis simply counts the number of diseases to evaluate the prevalence of multimorbidity. Such count based methods are a useful proxy for the complexity of healthcare needs in a population. However, the precise combination of conditions that tend to co-occur or cluster together is clinically important. In our review, we use existing systematic reviews on multimorbidity to anchor our discussion. We start with the methods commonly used to analyze multimorbidity, then discuss the disease relatedness patterns commonly found in the literature, and highlight a few notable studies. 

\subsection{Methods used to identify multimorbidity patterns}

Several systematic reviews have surveyed the methods used to identify multimorbidity patterns. Across the 14 studies reviewed by \textcite{prados2014multimorbidity}, cluster analysis was the most common technique, followed by exploratory factor analysis (EFA), the observed-to-expected ratio and multiple correspondence analysis. The review by \textcite{ng2018patterns}, which focuses specifically on analytical methods for disease cluster identification, found that 37 of the 41 studies it identified (90\%) used either factor analysis or hierarchical clustering, and \textcite{busija2019replicable} also include latent class analysis (LCA).

EFA studies often used tetrachoric correlations, a metric that measures how often two binary variables (diseases) occur together or not. Cluster analysis (CA) mostly used Yule’s Q (similar to tetrachoric) along with average linkage methods from hierarchical clustering (\cite{ng2018patterns, busija2019replicable}). Of note, the widely-used $k$-means method is rarely used in identifying multimorbidity patterns (\cite{busija2019replicable}). This is because $k$-means relies on Euclidean distance measures as a similarity metric, and while this is appropriate for continuous features, multimorbidity datasets involve binary features -- i.e., they are based on the presence or absence of each disease. 

\textcite{busija2019replicable} also took the diseases in the multimorbidity profiles they identified and conducted a meta-analysis using a method called multidimensional scaling (MDS). The MDS approach is used to visualize co-occurrence gradients, such as disease severity or well-known comorbidity groupings like metabolic or mental health profiles. This was achieved by evaluating model fit (\cite{kruskal1964multidimensional}) using Kruskal’s Stress values and interpretability of the solutions (\cite{borg2017choice}). However, it should be noted that the MDS method does not actually generate clusters. Instead, it uses dimensionality reduction techniques to illustrate how close different diseases might be to one another.

Within the domain of association rule mining, the FP-growth algorithm (\cite{han2000mining}) has become a landmark standard for extracting frequent patterns without the need for candidate generation. It is highly effective for discovering the most frequently observed co-occurrences in large datasets. Complementing this, hierarchical clustering—often using Ward’s linkage—is a well-established technique that groups diseases based on a proximity metric (e.g., \cite{foguet2015multimorbidity}). These methods provide essential benchmarks for pattern discovery; however, they often prioritize high-prevalence combinations.

Further, by applying the methods identified on the Australian National Health Survey (NHS) dataset, \textcite{ng2018patterns} find that ``even the same analytical method with different criteria for the proximity measure or the distance between clusters may obtain various results''. They also recommend that ``a detailed description of the methodology used when reporting and summarizing patterns of multimorbidity'' is crucial. Accordingly, in this paper, we provide a clear and transparent description of all the steps in our methodology, including the dataset, coding of the diagnoses, the similarity metric used, and steps in the extraction of clusters.

\subsection{Findings from systematic reviews}

The reviews also report on the disease-relatedness patterns that those methods found. The review by \textcite{prados2014multimorbidity} identifies key multimorbidity patterns from 14 studies in their review. Key clusters include diseases from cardiovascular and metabolic categories; mental health problems; and musculoskeletal problems. 

A more recent systematic review is provided by \textcite{busija2019replicable}. The authors synthesize multimorbidity profiles from 51 different articles, suggesting an increased interest in the field. Some analyzed clusters of patients while others identified clusters of diseases. Regardless, the studies reported groups of diseases, and \textcite{busija2019replicable} report a total of 407 multimorbidity profiles. 

Among the multimorbidity profiles identified, recurring disease clusters often involved mental health and cardio-metabolic conditions, suggesting true clinical relevance rather than statistical noise or error. While mental disorders are common, severe cases tend to be clustered within a small group of individuals who experience high levels of comorbidity (\cite{kessler2005prevalence}), and \textcite{andrews2002deconstructing} argue that, for clinical purposes, current comorbidity is best reduced to a principal disorder plus subsidiary disorders. Another well known connection is that the likelihood of developing cardiometabolic multimorbidity rises with higher BMI (\cite{kivimaki2017overweight}). Also, in a cohort of one million Chinese adults, cardiometabolic multimorbidity
more than doubled in prevalence over five years and was associated with a substantially elevated
mortality risk \parencite{zhang2019multimorbidity}. The consistency of these patterns across statistical methods and populations highlights shared underlying factors that could support targeted prevention strategies.

\subsection{Notable individual studies}

While the systematic reviews cover a wide range of studies, we now summarize notable examples that have further advanced the understanding of multimorbidity by identifying high-impact clusters and their implications for care delivery and patient outcomes. For instance, \textcite{soley2021identifying} conducted a longitudinal study in a multi-ethnic urban UK population, finding that patients in clusters involving substance dependence, HIV, and mental health disorders had the highest increases in primary care usage as additional conditions accumulated, and that these effects differed across ethnic groups. Similarly, \textcite{lenzi2016burden} examined multimorbidity patterns across age, gender, and citizenship in a large Italian population, revealing that older adults and native citizens had a higher burden of multimorbidity and that five distinct patterns could be identified through factor analysis. Complementing these epidemiological insights, \textcite{vaghfi2025identifying} developed an optimization framework to detect the most lethal combinations of comorbid diseases using electronic health records, offering a scalable method for mortality risk stratification in healthcare systems. Additionally, \textcite{goldstein2009comorbidity} used cluster analysis to uncover unique comorbidity profiles among homeless veterans, demonstrating how psychiatric and medical conditions frequently co-occur in this vulnerable group and vary by demographic factors.

Table~\ref{tab:multimorbidity_main} in the Appendix summarizes these and the other key studies in the multimorbidity profiling
literature, with the population, methodology and principal contribution of each.

\subsection{Data-driven clustering versus Medical Large Language Models}
Large language models (LLMs) can now answer free-text medical questions, and some have
reached passing-level performance on US medical licensing examinations
\parencite{thirunavukarasu2023large}. OpenEvidence is one example of a medical LLM that has been evaluated in US primary care \parencite{hurt2025use}. However, such models are judged by how closely their answers agree with the prevailing scientific consensus, and that consensus may be lacking for some subpopulations \parencite{singhal2023large}. In contrast, data-driven clustering approaches, such as the new clustering algorithm we propose, are designed to discover as-yet-unstudied associations directly from patient-level data. Clinical large language models, however, are a useful complement to our algorithm, since they can be used to check whether candidate associations extracted by the clustering methods have clinical support in the medical literature.

\subsection{Our Approach: RLSCA and Summary of Contributions}

While the approaches used in the literature vary by the clustering method and similarity
metric, they are by and large off-the-shelf approaches applied to multimorbidity. In
contrast, we propose a new methodology -- Randomized Length Sequential Clustering Algorithm
(RLSCA) -- that is highly customized to the characteristics of the multimorbidity problem.
RLSCA explores clusters of 3--8 diseases. Rather than restrict ourselves to a final
clustering solution as other methods do, we use randomized cluster lengths to create a
\textit{database} of most related disease subsets stratified by demographic groups, which
epidemiologists and clinicians can query and explore. Diseases are linked through lift, a
measure from the association mining literature: the ratio of the joint occurrence estimate
of two diseases to the product of the occurrence of the individual diseases. We allow lift
values to be included in our model only if joint support of the two diseases observed in
the dataset is greater than some user-specified threshold. Clustering methods in the past
have used metrics such as tetrachoric correlation; however, the inclusion of a support
threshold, which allows users to trade off statistical reliability with discovery of more
rare patterns, has not, to our knowledge, been used in multimorbidity clustering. Finally,
our model can be run separately for any demographic subgroup (this paper uses age and sex), enabling finer subgroup
profiling.

\section{Methodology}
\subsection{Data Acquisition and Imputation}

We begin by acquiring data from the Medical Expenditure Panel Survey (MEPS; \cite{ahrq_meps}) covering the years 2016 to 2019. MEPS public-use files are de-identified and released publicly by the Agency for Healthcare Research and Quality (AHRQ), and participation in the survey is voluntary; because this study is a secondary analysis of these public, de-identified files, it did not require review by an Institutional Review Board. Specifically, we use the Medical Conditions File and the Full Year Consolidated Data File. The Medical Conditions File contains patient records with a unique anonymous identifier (DUPERSID) and their associated medical conditions in ICD-10 codes. The Full Year Consolidated Data File provides demographic and weighting information for the U.S. population, including age, sex, ethnicity, and the weight each individual represents in the population—that is, the number of people in the U.S. population that the individual represents. The sum of weights of all the individuals equals the size of the U.S. population. 

ICD-10 codes are alphanumeric and provide detailed descriptions of medical conditions. Table~\ref{tab:icd_ccc} presents
examples of ICD-10 codes from 2019. Each ICD-10 code maps to a broader category of diseases, which we refer to in this study as Clinical Classifications Code (CCC), also known as Clinical Classifications Software (CCS) categories. These were defined by AHRQ. We convert ICD-10 codes to CCCs because CCCs strike a balance: they are descriptive enough for our purposes but not so granular that the data becomes unmanageable.

Our initial goal is to construct a dataset where each row represents a surveyed individual—or more accurately, a weighted group of patients in the U.S.—and each column corresponds to a CCC (i.e., a disease group). The cell values are binary indicators showing whether a patient has a particular disease (1) or not (0). Table~\ref{tab:dataset} presents a
small sample of this final dataset, along with examples of the diseases included.  For each individual, we retain associated metadata including DUPERSID, age, sex, and population weight.

\begin{table}[htbp]
\centering
\caption{Disease Description with ICD-10 Codes and CCC}
\label{tab:icd_ccc}
\begin{tabular}{|p{7cm}|c|c|c|}
\hline
\textbf{Disease Description} & \textbf{ICD-10 Code} & \textbf{First 3 Digits} & \textbf{CCC} \\
\hline
Type 1 diabetes mellitus without complications & E109 & E10 & 49 \\
Essential (primary) hypertension & I10 & I10 & 98 \\
Pure hypercholesterolemia & E780 & E78 & 53 \\
\hline
\end{tabular}
\end{table}

\begin{table}[!htbp]
\centering
\caption{Final Dataset Format (2019 as an Example). Numbers in parentheses are CCC codes; Table~\ref{bodysystems} lists them by body system.}
\label{tab:dataset}
\begin{tabular}{|l|c|c|c|}
\hline
\textbf{Patient ID} & \textbf{Hypertension (98)} & \textbf{Hyperlipidemia (53)} & \textbf{Diabetes (49)} \\
\hline
2320005102 & 1 & 0 & 0 \\
2320006101 & 0 & 0 & 0 \\
2320006102 & 0 & 0 & 0 \\
2320006103 & 0 & 0 & 0 \\
2320008102 & 0 & 0 & 0 \\
2320012102 & 1 & 1 & 1 \\
2320013101 & 1 & 1 & 0 \\
\hline
\end{tabular}
\vspace{1em}

\textbf{Total Number of Patients in 2019 = 28512} \\
\textbf{Total Number of CCC codes = 193}
\end{table}

One caveat is that while full ICD-10-CM codes have up to seven characters, we only have access to the first three. Fortunately, the first three digits are, for the most part, enough for us to derive the broader CCC-based disease category. However, in a minority of cases, an ICD-10 code presented in the data might result in multiple
CCCs, as shown in Table~\ref{tab:ccc_impute} in the Appendix, and it might not be clear which to select. To counter this problem, we use MEPS data from 2010 to 2015 for imputing the broader disease category represented by CCC. This is because, prior to 2016, the survey reported the precise CCC(s) each person had. (MEPS had a system of using ICD-9 code and CCC codes until 2015, and ICD-10 codes replaced them starting in 2016.)

In other words, our imputation uses historical data on CCC prevalence specific to an age and gender subgroup to derive the CCC from the 3-digit ICD-10. It should be noted though that the number of times imputation was needed was low. The full procedure, with an example, is given in the Appendix. 

Additionally, we removed the unclassified diseases and symptoms from our final data set since these CCC codes do not refer to specific disease categories. We have final data sets from 2016 to 2019.

\subsection{Clustering via Mixed Integer Programming}

At the heart of our clustering algorithm is a graph partitioning problem, which we describe in this section. We begin with some preliminaries. The input to our clustering model is a binary dataset of patients and diseases. Let $k=1,2,\cdots,N$ denote the number of patients in the dataset in a particular year of MEPS. In MEPS, patients are also assigned  a weight $w_k$, so that  $\sum_{k}^{N} w_{k}$ denotes the size of the US population. Such weights are standard practice in national surveys where a much smaller subsample is used to make inferences about the full population. In MEPS 2019, for instance, 28,512 individuals were surveyed, and their combined weight adds up to approximately $327$ million, the estimate of the US population in 2019. 

In our analysis, we further adjust the data for age and gender. Let $g=1,2,..,G$ denote the set of age and gender based demographic groups, and let $N_g$ denote the number of patients in demographic group $g$. In our results, we use age groups in multiples of 20 ($0-20$, $20-40$ , $40-60$) and $60+$ in combination with gender, as reported by the surveyed individuals (Male or Female). While we use age and gender demographic groups, our method is more general in that other factors such as income ranges (that indicate socioeconomic status), insurance type (private insurance vs. government-funded insurance such as Medicaid), geography (urban vs. rural) can also be used. 

\subsubsection {Graphical Representation and the Concept of Pairwise Lift}

Let $I$ denote the set of all diseases (i.e., the set of all CCC codes (also known as CCS categories)) in the dataset and let $D = |I|$ denote the total number of diseases. For a given age and gender demographic consisting of $N_g$ individuals, the binary parameter $z_{i,k}, i \in I$ indicates whether disease $i$ is present in individual $k$ ($1$) or not ($0$). 

Next, we define an undirected graph in which each node represents a disease and an edge links a pair $(i,j)$ of diseases, where $i \in I$ and $j \in I$. Each edge is associated with a value that indicates how closely the diseases are linked -- that is, a measure that quantifies the correlation between a pair of diseases. While any measure of correlation can be used, in this paper, we use a simple and intuitive measure called \emph{lift}, which we borrow from the association rule mining literature. The lift $L_{i,j}$ for a pair of diseases $(i,j)$ is defined as follows: 

\begin{equation}
L_{i,j} = \frac{\widehat{P}(i \cap j)}{\widehat{P}(i) \widehat{P}(j)} 
\end{equation}
Thus, the lift of a pair of diseases is the ratio of the joint probability estimate of having both diseases to the product of the estimated probability of having each of the diseases. The ratio is symmetric for any disease pair $(i,j)$, that is  $L_{i,j} = L_{j,i}$. For a particular age and gender demographic $g$, the estimated probabilities can be easily calculated from the dataset as follows: 
\begin{equation}
\widehat{P}(i ) = \frac{\sum_{k=1}^{N_g} w_{k}(z_{ik})}{\sum_{k=1}^{N_{g}} w_{k}}, \text{and     }         \widehat{P}(j ) = \frac{\sum_{k=1}^{N_g} w_{k}(z_{jk})}{\sum_{k=1}^{N_{g}} w_{k}}  
\end{equation}
\begin{equation}
\widehat{P}(i \cap j) = \frac{\sum_{k=1}^{N_g} w_{k}(z_{ik}*z_{jk})}{\sum_{k=1}^{N_{g}} w_{k}} 
\end{equation}
The intuition underlying pairwise lift is as follows. If the two diseases are independent of each other, the ratio will be approximately equal to 1, since independence implies that the joint probability estimate will approximately equal the product of the marginals. An example of independent diseases would be a sprain (that occurs due to an accident), and high cholesterol; there is no reason to believe that the two diseases influence each other. If the two diseases are negatively associated with each other (i.e., the presence of one decreases the chance of having the other), the joint probability estimate will be low in relation to the product of the marginals, and therefore the ratio will be less than 1. For instance, many studies have established that the risk of rheumatoid arthritis is reduced in those who have schizophrenia (\cite{oken1999issue}).  

Finally, if the two diseases are positively associated with each other (i.e., if the presence of one increases the risk of the other), the joint probability estimate will be higher than the product of the marginals -- i.e., the diseases co-occur more than the product of the marginals would indicate. Thus, the ratio that defines pairwise lift will be higher than 1. It is this case that is of most interest to us, since we are trying to cluster diseases that are strongly associated/related with each other. A well-known example of positive association is between high blood pressure (also known as hypertension) and risk of heart attack (known as myocardial infarction).  

We include the lift value for a pair of diseases $(i,j)$ only if the support for their joint occurrence is greater than a predetermined threshold $T$. Formally, we calculate the lift of $(i,j)$ only if the weighted joint prevalence
$\widehat{P}(i \cap j)$ defined above exceeds $T$; otherwise we set $L_{i,j} = 0$.  
Larger values of $T$ allow only the most established and commonly occurring disease co-occurrences to have a non-zero lift value. However, by decreasing $T$, clinicians and epidemiologists can begin to include less frequently occurring pairs that may nevertheless have a high pairwise lift. Thus at lower values of $T$, clinicians and epidemiologists are more likely to include less common but highly related disease pairs in the clustering model.

\subsubsection {Mixed Integer Programming Model}
   
Once the graph is constructed and the pairwise lifts are calculated, our goal is to select a subset/cluster of $R$ diseases from the full set of $D$ diseases, such that \textit{the sum of the pairwise lift values of the diseases in the subset is maximized}. We require that each disease in the selected cluster must have a non-zero, statistically reliable pairwise lift with at least one of the other diseases in the subset. Thus, the model starts with pairwise lift values, but begins to group larger collections of diseases. It thereby enables clinicians and epidemiologists to discover connections between diseases that have not been established yet.

\begin{figure}[!htbp]
\centering
\begin{minipage}{0.5\textwidth}
\centering
\begin{tikzpicture}[scale=1.2, every node/.style={circle, draw, minimum size=1.2cm, inner sep=0pt}, node distance=2cm]

\node (49) at (0,0) {49};
\node (98) at (1,-1.5) {98};
\node (53) at (1,1.5) {53};
\node (95) at (3,-1.5) {95};
\node (50) at (3,1.5) {50};

\draw (49) -- (53);
\draw (49) -- (98);
\draw (49) -- (95);
\draw (49) -- (50);
\draw (98) -- (53);
\draw (98) -- (95);
\draw (98) -- (50);
\draw (53) -- (50);
\draw (53) -- (95);
\draw (95) -- (50);

\draw[dashed, red, thick] (2,-2) -- (2,2);
\draw[->, thick] (3.5,0) -- (5.5,0);

\begin{scope}[xshift=6.5cm]
\node (49c) at (0,0) {49};
\node (98c) at (2,0) {98};
\node (53c) at (1,1.8) {53};
\draw (49c) -- (98c);
\draw (49c) -- (53c);
\draw (98c) -- (53c);
\end{scope}

\end{tikzpicture}
\end{minipage}%
\hspace{3.5cm} 
\begin{minipage}{0.25\textwidth}
\footnotesize
\begin{flushleft}
\textbf{Disease legend}\\[-0.3em]
\setlength\leftmargini{1.3em} 
\setlength\itemindent{0em}
\begin{itemize}
  \item 49: Diabetes
  \item 53: Hyperlipidemia
  \item 98: Hypertension
  \item 50: Diabetes (with complications)
  \item 95: Other nervous system disorders
\end{itemize}
\end{flushleft}
\end{minipage}

\caption{Example of a fully connected disease clique (left) and the selected cluster (right) with the corresponding disease legend. The numbers in the nodes are CCC codes; Table~\ref{bodysystems} lists them by body system.}
\label{fig:cluster_example}
\end{figure}
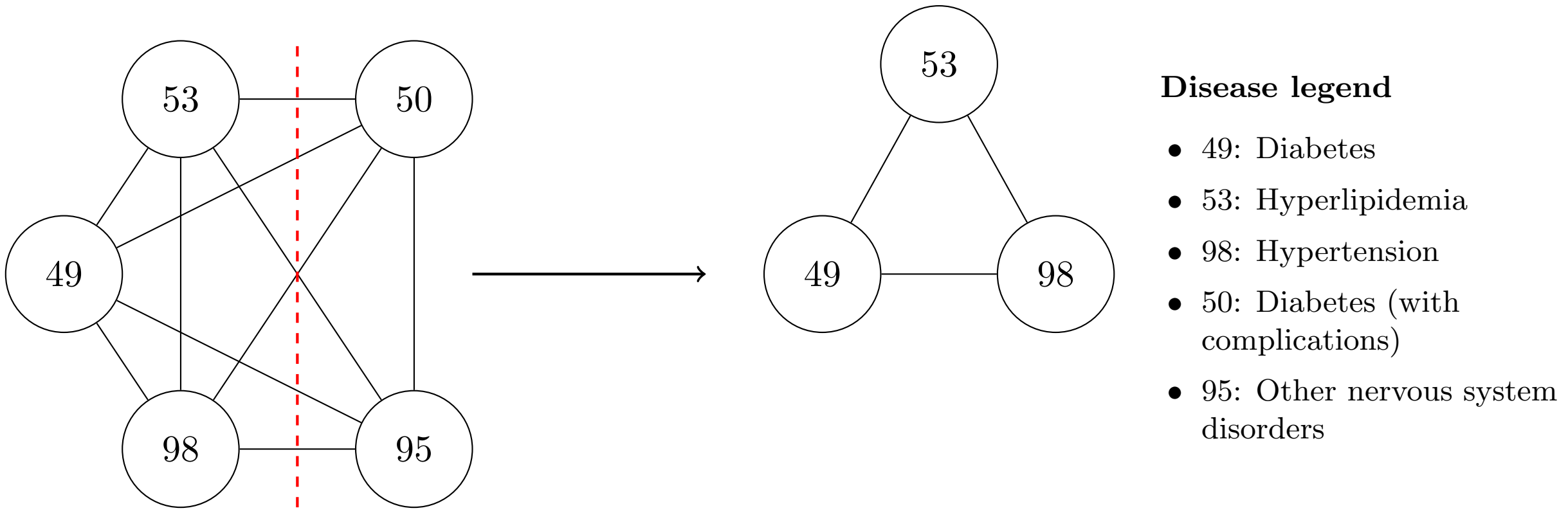

We illustrate this basic idea with Figure~\ref{fig:cluster_example} using a simple graph consisting of five diseases. The goal in this case is to choose a subset of three diseases that have the greatest pairwise sum of lift values. 

We now formalize the ideas described above into a mixed-integer programming (MIP) model.  In Table~\ref{tab:notation}, we present the notation used in the MIP model, including the sets, parameters, and decision variables along with their corresponding descriptions. The set represents the nodes and edges of the disease network, while the parameters and decision variables define the model’s structure and decision logic.

\begin{table}[!htbp]
\caption{Sets, Parameters, and Variables}
\label{tab:notation}
\vspace{0.3cm}
\centering
\renewcommand{\arraystretch}{1.2}
\begin{tabular}{p{3cm} p{12cm}}
\toprule
\textbf{Symbol} & \hfill \textbf{Description} \\
\midrule

\multicolumn{2}{l}{\textbf{Sets}} \\[2pt]
$i, j \in I$ & \hfill Nodes (diseases) \\[6pt]

\multicolumn{2}{l}{\textbf{Parameters}} \\[2pt]
$\epsilon$ & \hfill Small positive constant ($10^{-5}$ in our implementation) \\
$\Lambda$ & \hfill Large constant ($10^{4}$ in our implementation) \\
$R$ & \hfill The length chosen for the cluster \\
$L_{i,j}$ & \hfill Lift value for $(i, j)$ pair \\[6pt]

\multicolumn{2}{l}{\textbf{Variables}} \\[2pt]
$X_i \in \{0,1\}$ & \hfill Binary variable for choosing node $i$ for the cluster \\
$Y_{i,j} \in \{0,1\}$ & \hfill Binary variable indicating whether pair $(i, j)$ is in the cluster \\
\bottomrule
\end{tabular}
\end{table}

Next, we present the objective function and constraints defined in Equations~\eqref{eq:objective}–\eqref{const:binary}.

\begin{align}
\max \quad Z &= \sum_{i} \sum_{j>i} Y_{i,j} L_{i,j} 
\label{eq:objective} \\
\text{s.t.} \quad
& \sum_{j} Y_{i,j} L_{i,j} \geq \epsilon + \Lambda (X_i - 1), && \forall i \in I 
\label{const:activation} \\
& \sum_{i} X_i \leq R, 
\label{const:resource} \\
& Y_{i,j} \leq X_i, && \forall i,j \in I
\label{const:upperlink} \\
& Y_{i,j} \leq X_j, && \forall i,j \in I
\label{const:upperlink2} \\
& Y_{i,j} \geq X_i + X_j - 1, && \forall i,j \in I 
\label{const:lowerlink} \\
& X_i, Y_{i,j} \in \{0, 1\}, && \forall i,j \in I
\label{const:binary}
\end{align}

As shown in Equation~\eqref{eq:objective}, the model maximizes the total lift values of all disease pairs within the selected cluster, thereby strengthening the internal connections among its members. 

Constraint~\eqref{const:activation} ensures that if a disease (node) is included in the cluster (\(X_i = 1\)), it must have a non-zero lift with at least one other disease in that cluster. 

Constraint~\eqref{const:resource} limits the number of diseases that can be included in a cluster based on the randomly generated upper bound \(R\). This constraint sets the cluster length to a maximum of \(R\). Because the objective seeks to maximize intra-cluster lift values, the constraint typically binds as an equality. However, in cases where fewer eligible diseases are available, this constraint prevents infeasibility by allowing the cluster to include as many valid nodes as possible. 

Constraints~\eqref{const:upperlink}, \eqref{const:upperlink2} and~\eqref{const:lowerlink} together define the binary decision variable \(Y_{i,j}\), which indicates whether a pair of diseases \(i\) and \(j\) belong to the same cluster (\(Y_{i,j} = 1\)). Finally, Constraint~\eqref{const:binary} specifies the binary nature of all decision variables in the model.

\subsection{Randomized Length Sequential Clustering Algorithm (RLSCA)}

We embed the MIP described above within a randomized length sequential clustering algorithm (RLSCA), shown in Algorithm \ref{alg:RLSCA}. The input to the algorithm is the full graph $G$ of diseases with the pairwise lift values serving as weights on the edges, and an acceptable threshold, $T$, for calculation of pairwise lift. All disease pairs that do not meet this threshold are assigned a lift of 0.  Another input is a limit $nd \le D$ on the number of diseases that will be clustered. It should be noted that the double-indexing notation used in Algorithm~\ref{alg:RLSCA} is intended solely to track clusters within their specific iterations. For all subsequent outputs---whether referencing the comprehensive cluster database or the final disease groupings---a simplified single-index notation is adopted for clarity and ease of interpretation.

\begin{algorithm}
\caption{Randomized Length Sequential Clustering Algorithm}
\label{alg:RLSCA}
\begin{algorithmic}[1]
\State \textbf{Step 1:} Set minimum support threshold \( T \) and eliminate disease pairs \( (i,j) \) below this threshold by setting their lift values \( L(i,j) = 0 \). \hfill 
\For{$m = 1$ to $M$}  \hfill \textit{(Iterations \( M \))}
    \State \( k = 1 \)
    \Repeat
        \State Draw a random cluster length \( R \) between \( lb \) and \( ub \).
        \State Find a cluster \( C_{km} \) using MIP, maximizing total lift. \hfill \textit{(Cluster \( C_{km} \))}
        \State Remove all diseases in \( C_{km} \) from future selection: 
        \State \hspace{1cm} Add the constraint \( X_i = 0 \quad \forall i \in C_{km} \) to the MIP model. \hfill 
        \State\( k++ \)
    \Until{Total diseases selected \( \geq nd \), or no cluster can be formed.} \hfill \textit{(Number of Diseases \( nd \))}
    \State \textbf{Store clusters:} \( C_{1m}, C_{2m}, \ldots, C_{km} \rightarrow \text{Database} \). \hfill \textit{(Storage Step)}
    \State \textbf{Reset all the clusters:}
    \State \hspace{1cm} Remove all constraints \( X_i = 0 \quad \forall i \in C_{1m}, C_{2m}, \ldots, C_{km} \). \hfill \textit{(Deactivation Step)}
    \State \hspace{1cm} Clear cluster assignments \( C_{1m}, C_{2m}, \ldots, C_{km} \). \hfill \textit{(Reset Step)}
    
\EndFor
\State \textbf{Output:} Database of clusters obtained from all iterations.
\end{algorithmic}
\end{algorithm}

The algorithm runs for $M$ iterations and each iteration consists of a sequence of steps. At the beginning of each iteration, we have the $G$ and the $L_{i,j}$ for all $(i,j) \in I$. In each step of an iteration, we use the MIP to extract the $R$ most related diseases in the graph, where  $lb \le R \le ub$ and $R$ follows a discrete uniform distribution in the range $[lb,ub]$, drawn anew at each step. Since most multimorbid patients (about 61\% in each year of our data) have between $3$ and $8$ diseases, we set $lb = 3$ and $ub = 8$.  We note that randomization is an important component of the algorithm, since it allows different subsets of diseases to cluster together across iterations. 

Once a cluster of $R$ diseases is chosen, those diseases will no longer be chosen in future steps within the iteration.  This is ensured by setting the decision variable $X_i = 0$ corresponding to each disease in the cluster chosen. Thus, when an MIP is executed in a future step, diseases already chosen in prior clusters are not chosen again. The process of selecting clusters sequentially continues until we reach a step in which the total number of diseases clustered in previous steps reaches a user-specified number $nd \le D$, or until no further cluster can be formed because no two remaining diseases have a non-zero lift.  All the clusters identified in each iteration are added to a database; the next iteration starts once again with the original graph $G$ and $L_{i,j}$ for all $(i,j) \in I$. At the end of the algorithm, we have a database of clusters from all the $M$ iterations. 

\noindent \textbf{Sample Results and Clinical Validity}: As an example, Table~\ref{tab:rlsca_results} shows some sample clusters identified by RLSCA, using a  support threshold  of $T = 0.018$ for all individuals surveyed in MEPS $2019$. The RLSCA uses this threshold value -- which translates to a prevalence of approximately 6 million in the US population -- to determine whether lift between a pair of diseases is statistically reliable to include in our model. Only pairs of diseases that have a joint prevalence higher than this threshold will have a non-zero lift value in our model; all other pairs will have a pairwise lift of 0. 

\begin{table}[!htbp]
\centering
\caption{Diverse disease clusters (all ages) with disease names, diversity values, and frequencies using support of 0.018. Numbers in parentheses are CCC codes; Table~\ref{bodysystems} lists them by body system.}
\begin{tabular}{|p{11cm}|c|c|}
\hline
\textbf{Cluster (Disease Name + Code)} & \textbf{Diversity} & \textbf{Frequency} \\
\hline

Anxiety disorders (651), Mood disorders (657) & 1 & 53 \\
\hline

DiabMel no c (49), Hyperlipidem (53), Hypertension (98), Esophgeal dx (138), Osteoarthritis (203)& 4 & 20 \\
\hline

Thyroid dsor (48), DiabMel no c (49), Hyperlipidem (53), Hypertension (98), Esophgeal dx (138), Osteoarthritis (203), Anxiety disorders (651), Mood disorders (657) & 5 & 18 \\
\hline

\end{tabular}
\label{tab:rlsca_results}
\end{table}

A high threshold value of $0.018$ is expected to reveal common conditions that are well known to be related or positively associated with each other together. This allows us to validate the clusters produced by RLSCA. In Table~\ref{tab:rlsca_results} we observe three different example clusters of size $2$, $5$ and $8$ that are identified frequently during 100 iterations of the RLSCA. The Frequency column gives the number of the 100 randomized iterations in which the algorithm returned that cluster, a measure of how consistently the optimizer rediscovers it, not of how many patients have the combination.

We additionally quantify the number of body systems represented in a cluster using a metric that we call \textit{Diversity}. Specifically, the first cluster consists of two mental health conditions: Anxiety Disorders and Mood Disorders (better known as Depression), which was picked in 53 of the 100 iterations of the algorithm. This cluster has a diversity of 1, since both diseases belong to the mental health body system. 

The second cluster of five diseases spans four different body systems: Diabetes and Hyperlipidemia (Endocrine and Metabolic Systems); Hypertension (Circulatory System); Esophageal Disorders (Digestive System); and Osteoarthritis (Musculoskeletal System).  Modern medicine often operates under specialty silos which are narrowly defined by individual organs or body systems -- for instance, cardiology (heart), nephrology (kidneys), neurology (nervous system), etc. By identifying clusters that belong to multiple body systems, our approach promotes a more holistic and integrated look at how diseases relate to each other.  A larger cluster like this that is chosen frequently  provides a stronger signal that the diseases may be physiologically connected. For instance, \textcite{reaven1988role} describes insulin resistance as a common link between
hypertension, type 2 diabetes, and abnormal blood lipids, a trio of chronic conditions well known to clinicians.  The inclusion of Osteoarthritis within metabolic clusters is consistent with the proposed systemic inflammatory link between metabolic syndrome and osteoarthritis (\cite{li2016metabolic}).

The last cluster in Table~\ref{tab:rlsca_results} involves 8 diseases, and was picked 18 times in the 100. It has a Diversity of 5. This cluster  combines diseases in the other two clusters but now also includes thyroid disorders, a different disease of the endocrine system.
Thyroid stimulating hormone has been shown to sit at the center of the
symptom network in first-episode major depressive disorder
\parencite{peng2023association}, consistent with a thyroid--mood
association.

These results at a high threshold value ($T = 0.018$) provide empirical validation of the RLSCA framework. In the Results section, we use lower threshold values to detect lesser known disease clusters in different age and gender demographic groups.

\subsection{Algorithm for Choosing Final Clusters}

The database is one of the outputs of our algorithm, and can be directly queried by physicians and epidemiologists. However, as observed in the previous section, clusters can have overlapping diseases. If a single final clustering solution with non-overlapping diseases is desired, then the approach outlined in Algorithm~\ref{alg:nonoverlap} can be used. 

The algorithm takes as input the top $n$ clusters in the database based on frequency. Next, the clusters are sorted first based on their length; ties are broken based on frequency. Starting with the first cluster in this list, the algorithm iterates through the list, progressively adding non-overlapping clusters. This yields a final set of non-overlapping clusters. There are other ways to create non-overlapping clusters -- for instance, clusters could be sorted based on diversity rather than length (length and diversity can be quite correlated, however), or clusters could be sorted purely based on their frequency. If the goal is discovery of new clinical relationships, sorting by length or diversity would be more effective. If the goal is to detect diseases that are more likely to occur together, sorting by frequency is a better choice. 

\begin{algorithm}
\caption{Selection of Final Non-Overlapping Clusters}
\label{alg:nonoverlap}
\begin{algorithmic}[1]
\State \textbf{Input:} The set of top $n$ clusters with the highest frequencies, denoted as $\mathcal{C} = \{C_1, C_2, \dots, C_n\}$.
\State \textbf{Step 1:} Sort the set $\mathcal{C}$ such that:
\State \quad (i) Cluster lengths follow $R_1 \geq R_2 \geq \dots \geq R_n$
\State \quad (ii) For equal cluster lengths $R_i = R_j$, sort such that $a_i > a_j$ (sort by decreasing frequency)
\State \textbf{Step 2:} Initialize an empty set $\mathcal{B} = \emptyset$ (Final selected clusters).

\For{$i = 1$ \textbf{to} $n$}
    \State Let $C_i$ be the current candidate cluster from the sorted set.
    \If{$C_i \cap (\cup_{C \in \mathcal{B}} C) = \emptyset$} \Comment{Verify no overlap with set $\mathcal{B}$}
        \State $\mathcal{B} \gets \mathcal{B} \cup \{C_i\}$ \Comment{Add cluster to final selection}
    \EndIf
\EndFor

\State \textbf{Output:} Final set of non-overlapping clusters $\mathcal{B}$.
\end{algorithmic}
\end{algorithm}

The results of applying Algorithm~\ref{alg:nonoverlap} to the MEPS 2019 database are presented in Section 5.3.

\subsection{RLSCA in Practice}

We envision two distinct uses for the RLSCA and the resulting database of disease clusters. First, it can be used by clinicians at the point of care. For instance, primary care physicians routinely see complex patients in their practices whose chronic condition combinations they have never encountered before (as we pointed out, this is due to the heterogeneity in condition combinations). By inputting their condition combinations along with demographic characteristics, the physicians can learn other diseases the patient is at risk of developing, and thereby suggest prevention and screening strategies. The database and cluster visualizations can be integrated into a large language model (LLM) trained on the medical literature. OpenEvidence is an example of such a platform. As of now, OpenEvidence is able to search and summarize the vast corpus of medical literature. Physicians can use it when advising patients on their conditions. Coupling our clustering algorithm to a medical LLM such as OpenEvidence would create a powerful diagnostic tool to evaluate the data-driven clusters we generate with the published evidence.   

The second application of our algorithm is to facilitate the process of epidemiological discovery. Specifically, the RLSCA is designed to identify lesser-known associations in the data that have not been investigated yet because of high heterogeneity and low prevalence. Used in this way, RLSCA is a hypothesis generator.  These associations can be used by epidemiologists and researchers to evaluate whether valid causal/physiological explanations might exist. Negative results that disprove spurious associations are equally useful.

\section{Alternative Clustering Approaches}

As mentioned earlier, RLSCA is highly customized to find clusters of related diseases. We now illustrate and adapt two other approaches for disease clustering.

The first approach is a commonly used data-mining methodology called \textit{frequent itemset mining}. It is also called \textit{frequent pattern discovery} and -- because of its use in the retail context -- \textit{market-basket analysis}. In the retail context, the method is used to find subsets of items that are purchased together among millions of customer transactions -- as an example, bread, cheese, and peanut butter may be a commonly purchased trio of items in a grocery store. In the multimorbidity context, the goal is to identify the frequently occurring subsets of diseases using computationally efficient methods such as the Apriori and FP-growth algorithms.  While the algorithm itself is considered a landmark advance in data mining and is extensively used, here we adapt it to disease clustering as a baseline for comparison.  

The second approach is the highly popular agglomerative hierarchical clustering, the most commonly used method to cluster diseases in the literature. As with frequent itemset mining, we adapt hierarchical clustering to identify disease clusters. 

\subsection{Frequent Itemset Mining via the Apriori and FP-Growth Algorithms}

The Apriori algorithm is a classic method used for mining frequent itemsets and generating association rules. Traditionally, it identifies patterns by examining combinations of items — diseases in our case — that frequently co-occur in real-world data. It works iteratively, starting from individual items and gradually building up to larger sets that meet a minimum support threshold. In this context, an ``itemset'' refers to a group of diseases that tend to appear together in patients, forming a potential multimorbidity cluster (\cite{agarwal1994fast}).

Formally, consider a dataset of $N_{g}$ patients from a specific age and gender group $g$ and $i=1,2,\cdots, D$ diseases. For each disease $i$, we define a corresponding random variable $Z_i$ such that $Z_i = 1$ if a patient has disease $i$ and $Z_i = 0$ otherwise. The realization of this random variable in the dataset is represented by a binary variable $z_{ki}$ which  indicates whether disease $i$ is present in patient $k$ (indicated with a 1) or not (indicated with a 0).  The goal is to identify all subsets of diseases $\mathcal{I} \subset \{1,\dots,D\}$ for which the joint probability $\Pr\left[\bigcap_{i \in \mathcal{I}} Z_i = 1\right]$—interpreted as the support—is greater than some predefined threshold $t$. This probability measures how often all diseases in the subset $\mathcal{I}$ co-occur across the dataset. As discussed earlier, each patient $k$ is associated with a weight $w_{k}$ which indicates the number of individuals in the US population that person represents. Thus, given a dataset of patients in age and gender group $g$, we can estimate the support for a subset of diseases, $\mathcal{I} $, as follows:

\begin{equation}
\Pr\left[\bigcap_{i \in \mathcal{I}} (Z_i = 1)\right] = \frac{1}{\sum_{k=1}^{N_{g}} w_{k}} \sum_{k=1}^{N_{g}}w_{k}\prod_{i \in \mathcal{I}} z_{ki}
\end{equation}

In frequent itemset mining, the goal is to discover all disease subsets that have a support of at least $t$. In other words, we identify and list each subset of diseases that co-occurs in at least a fraction $t$ of the weighted population of group $g$. This threshold helps filter out low-prevalence combinations and manage the computational complexity of evaluating the $2^D$ possible subsets of $D$ diseases.

The Apriori algorithm uses this principle efficiently. In the first pass, it identifies individual diseases that meet the support threshold. In the second iteration, the algorithm considers pairs of the surviving individual diseases and retains the pairs that cross the support threshold $t$. Similarly, each subsequent pass combines surviving subsets from the previous iteration into a higher order itemset with one additional disease and retains only those that cross the support threshold. This iterative process continues until no more frequent itemsets can be generated (\cite{agarwal1994fast}). The more recent FP-Growth Algorithm is an improvement over the Apriori algorithm and improves the computational efficiency of finding itemsets. It compresses the database into a structure called the FP-tree and mines it recursively. The output of both algorithms (Apriori and FP-Growth) is the same, but since FP-Growth is faster, we will use it for our analysis (\cite{han2000mining}).

In our implementation of frequent itemset mining, each row in the input dataset represents a group of $w_{k}$ patients. Expanding the data fully according to these weights would create nearly 327 million rows, making the FP-Growth computationally difficult to execute. To address this, we use the weights to probabilistically sample a new dataset with 1 million rows for each full survey year from 2016 to 2019, and then apply the FP-Growth algorithm.
The probability that each row will be sampled is $\frac{w_k}{\sum_k w_k}$. This ensures that observations with larger weights (i.e., representing a larger number of Americans) have proportionally higher chances of being selected. Once the probabilities are established, we repeatedly sample rows from the dataset (with replacement) according to this distribution until a total of one million records have been sampled. This approach preserves the weighted population structure of the original data while producing a computationally manageable dataset to which the FP-Growth algorithm can be applied.

Table~\ref{tab:fp_results} presents example FP-Growth results obtained from the 1-million rows sampled from the 2019 dataset. Here we present the most frequently occurring itemsets of size three and four. For instance, hypertension (98), diabetes (49) and abnormal cholesterol levels (53) is the most common triad of co-occurring diseases in the United States, affecting $3.1 \%$ of the population. The most frequently occurring quartet of co-occurring diseases consists of the triad mentioned above and osteoarthritis. 

\begin{table}[!htbp]
\centering
\caption{Top FP-Growth clusters (1M dataset) with support and diversity values. Numbers in parentheses are CCC codes; Table~\ref{bodysystems} lists them by body system.}
\begin{tabular}{|p{11cm}|c|c|}
\hline
\textbf{Cluster (Disease Name + Code)} & \textbf{Support} & \textbf{Diversity} \\
\hline

Hypertension (98), DiabMel no c (49), Hyperlipidem (53) & 0.0314 & 2 \\
\hline

Hypertension (98), Hyperlipidem (53), Osteoarthritis (203) & 0.0176 & 3 \\
\hline

Hypertension (98), Ot conn tiss (211), Hyperlipidem (53) & 0.0166 & 3 \\
\hline

Hypertension (98), DiabMel no c (49), Hyperlipidem (53), Osteoarthritis (203) & 0.0071 & 3 \\
\hline

\end{tabular}
\label{tab:fp_results}
\end{table}

Thus, frequent itemset mining systematically generates disease clusters based on support, and this in and of itself, is useful for clinicians and  epidemiologists. However, the key downside of FP-Growth is that it primarily returns a large number of highly repetitive clusters already known to occur in the real data, offering limited new insight.  Further, the algorithm typically struggles to produce lower-support itemsets that have a larger number of diseases, which are often more informative. This is due to the algorithm’s heavy computational demands. At the lowest support we could compute (0.0015), FP-Growth returned 1,429 itemsets from the one-million-row 2019 sample: 1,036 of three conditions, 367 of four and 26 of five, and every five-condition itemset contains hyperlipidemia (53) and hypertension (98). By contrast, RLSCA returns clusters of up to eight conditions (Section 5).

The RLSCA algorithm that we propose has a different goal: instead of only capturing frequent historical combinations, we aim to uncover potential new connections among diseases from different body systems -- combinations that may or may not have appeared together in the data (due to the sparsity  of highly heterogeneous combinations) but are statistically or clinically meaningful. This approach broadens the exploration beyond what is merely observed, helping to simulate and anticipate possible co-occurrence patterns based on underlying pairwise relationships. It allows for discovery that is more exploratory in nature, potentially revealing hidden structures in the data that traditional methods like Apriori and FP-Growth may miss.

\subsection{Hierarchical clustering and comparison to RLSCA}
\label{sec:hc_benchmark}

Hierarchical clustering is a traditional unsupervised learning technique that organizes data into an upside-down nested tree structure, also known as a  dendrogram, based on a defined similarity measure. In multimorbidity research, this method is often used to group diseases that frequently co-occur within a population.

The process typically follows an agglomerative (bottom-up) approach:
\begin{itemize}
    \item \textbf{Initialization:} Each individual disease is initially treated as a separate cluster.
    \item \textbf{Distance Computation:} The algorithm calculates the distance between all pairs of clusters using a distance measure. In the case of disease clusters, distance is based on a correlation measure (lift in our case) that quantifies how related a pair of diseases are.
    \item \textbf{Iterative Merging:} At each step, the two closest clusters are merged based on a linkage criterion---such as Ward's method, which minimizes the total within-cluster variance.
    \item \textbf{Termination and Partitioning:} This cycle repeats until all diseases are consolidated into a single hierarchy. To obtain a final partition, the researcher must ``cut'' the dendrogram at a specific height or designate a target number of clusters.
\end{itemize}

While hierarchical clustering is a valuable benchmark, its deterministic nature and lack of control over cluster cardinality often result in large and uninformative groupings. This is what motivated the randomized approach.

As a benchmark to our Randomized Length Sequential Clustering Algorithm (RLSCA), we implemented an agglomerative hierarchical clustering approach on the same 2019 MEPS dataset and under conditions that are as comparable as possible to our method.

\paragraph{Data and similarity structure.}
For the hierarchical clustering, we restricted attention to the 2019 data and used the same set of diseases used in our previous analysis (all CCS-based disease categories that enter the frequent itemset mining step). To make the comparison fair, we applied the minimum support threshold that RLSCA uses in Section 5, approximately $0.0018$ (corresponding to a prevalence of 600,000 in the US population). Only disease pairs with support above this threshold keep their lift value; for all other pairs it is set to zero.

Using the same pairwise lift values defined in Section 3.2.1, we constructed a lift matrix where each entry $\text{lift}_{ij}$ is the lift for the rule ``disease $i$ $\rightarrow$ disease $j$''. Because hierarchical clustering operates on an undirected similarity (or distance) structure, we symmetrized this matrix by averaging the two directions (lift is symmetric by construction; this step is included so the framework also accepts asymmetric association measures):
\begin{equation}
\tilde{L}_{ij} = \frac{\text{lift}_{ij} + \text{lift}_{ji}}{2}, 
\qquad 
\tilde{L}_{ii} = 0.
\end{equation}

Here $\tilde{L}_{ij}$ measures the strength of association between diseases $i$ and $j$; higher values indicate stronger co-occurrence relative to independence.

\paragraph{Distance transformation and linkage method.}
Hierarchical clustering requires a distance matrix rather than a similarity matrix. To keep the scale of the distances as close as possible to the original lift values, we defined
\begin{equation}
d_{ij} = L_{\max} - \tilde{L}_{ij}.
\end{equation}

where $L_{\max} = \max_{i,j} \tilde{L}_{ij}$. With this transformation, pairs of diseases with high lift (strong association) have small distance, and pairs with zero lift have distance approximately equal to $L_{\max}$. The diagonal is set to zero, $d_{ii} = 0$, and the resulting symmetric distance matrix $\Delta = (d_{ij})$ is passed to a standard hierarchical clustering routine.

We experimented with several linkage functions (average, complete, single, and Ward). In practice, most linkage types either produced one giant cluster with many very weakly connected diseases or a very unbalanced structure dominated by a small number of large clusters. Ward's method, which minimizes the increase in within-cluster variance at each merge, produced the most balanced
partition of the four we tried. Therefore, all hierarchical results reported in this section are based on Ward linkage applied to the lift-based distance matrix $\Delta$.

Once the dendrogram is computed, one can either cut the tree at a specified distance threshold or request up to a given maximum number of clusters, but there is no way to constrain the length of each resulting disease cluster.  In our experiments, feeding the full set of diseases to Ward clustering produced many clusters with extremely long lengths (dozens of diseases in a single group). Such clusters are difficult to interpret clinically and not directly suitable for our goal of discovering compact ``cliques'' of 3--8 diseases (recall that most multimorbid patients have between 3 and 8 conditions). 

\begin{figure}[!htbp]
    \centering
    \includegraphics[width=0.95\linewidth]{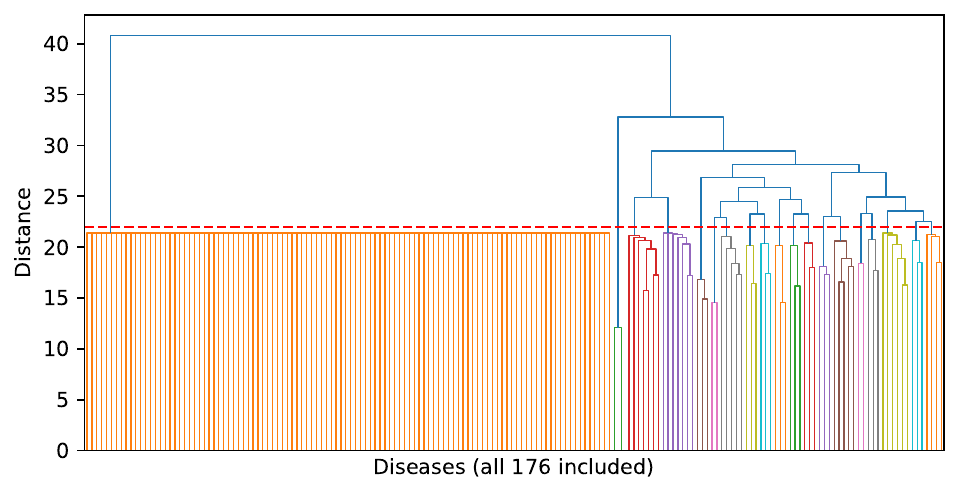}
    \caption{Ward hierarchical clustering dendrogram based on the 2019 lift-based distance matrix, with a dashed horizontal cut at distance 22 (19 clusters). All 176 diseases are included; disease pairs below the support threshold of 0.0018 have their lift set to zero.}
    \label{fig:ward_tree_2019}
\end{figure}

Figure~\ref{fig:ward_tree_2019} presents the full Ward hierarchical clustering dendrogram based on the 2019 lift-based distance matrix. Because all diseases are included, the tree is dense and visually difficult to interpret. The horizontal cut is shown only as an example, highlighting how a single cut partitions the diseases while also illustrating a key limitation of hierarchical clustering: the method cannot enforce cluster-size ranges and can produce very large, uninformative clusters (at the cut shown, one cluster holds 108 of the 176 diseases).

A second disadvantage relative to the RLSCA is that, for a fixed dataset, distance definition, and linkage method, the hierarchical clustering solution is fully deterministic: running it from scratch with the same inputs always yields the same dendrogram. The only flexibility comes from changing the support threshold (which we already vary in our own approach) or stratifying the data by subgroup (e.g., age, sex), not from the clustering procedure itself. Thus hierarchical clustering (as it is currently applied in most packages) is not well suited for discovering multiple clustering solutions.  

Despite these shortcomings, hierarchical clustering has two practical advantages: it is available in every standard statistical library, and it is well established in this literature.

\paragraph{Creating Stable Disease Clusters using Hierarchical Clustering.}
In what follows, we adapt hierarchical clustering to produce stable clusters by varying the number of clusters $k$ (details in the Appendix). 

Clusters that appear frequently across different cuts of the dendrogram can be interpreted as structurally stable in the hierarchical model, while clusters that appear only once are more sensitive to the choice of $k$.

Table~\ref{tab:hc_stability} lists four of the six clusters that appear in more than ten distinct partitions (the most frequent one and the three that span the most body systems), excluding the single large residual group containing every disease not otherwise clustered. These are the most stable outputs of hierarchical clustering under this procedure. 

\begin{table}[!htbp]
\centering
\caption{Most frequently appearing hierarchical clusters across $k = 3, \dots, 50$ using the \texttt{maxclust} criterion. The frequency column counts how many distinct partitions produced the same cluster; only clusters of manageable length are shown. Numbers in parentheses are CCC codes; Table~\ref{bodysystems} lists them by body system.}

\label{tab:hc_stability}
\begin{tabular}{|c|p{9cm}|c|c|}
\hline
\textbf{Cluster} & \textbf{Diseases (Name + Code)} & \textbf{Diversity} & \textbf{Frequency} \\
\hline

1 & Acute myocardial infarction (100), Coronary atherosclerosis and other heart disease (101), Transient cerebral ischemia (112) & 1 & 19 \\
\hline

2 & Acute cerebrovascular disease (109), Acute and unspecified renal failure (157), Hyperplasia of prostate (164), Cancer of prostate (29), Retinal detachments and defects (87), Peri-, endo-, and myocarditis; cardiomyopathy (97), Essential hypertension (98) & 4 & 12 \\
\hline

3 & Congestive heart failure; nonhypertensive (108), Aortic and peripheral arterial embolism or thrombosis (116), Ot dx kidney (161), Other male genital disorders (166), Diabetes mellitus without complication (49), Disorders of lipid metabolism (53), Gout and other crystal arthropathies (54) & 3 & 12 \\
\hline

4 & Chronic obstructive pulmonary disease and bronchiectasis (127), Asthma (128), Other lower respiratory disease (133), Other upper respiratory disease (134), Other gastrointestinal disorders (155), Other GU dx (163), Headache/mig (84), Other ear and sense organ disorders (94) & 4 & 11 \\
\hline

\end{tabular}
\end{table}

\section{RLSCA Results}

We now present a selection of results for the RLSCA for different age and gender subgroups in the US population.  Our analysis utilized the 2019 MEPS dataset, stratified into eight distinct demographic groups based on gender and age. Specifically, the 2019 dataset was divided into four age-based cohorts: 0--20, 20--40, 40--60, and 60+. We intentionally designed these intervals with overlapping boundaries at the 20, 40, and 60-year marks. We treat the boundary ages as transition points belonging to both adjacent groups. A twenty-year-old is plausibly described by either band, and including such individuals in both strata gives each stratum more data from which to estimate pairwise lift. The strata therefore overlap rather than partition the sample: the four age groups carry weighted totals of 87.2, 90.5, 86.1 and 74.1 million, summing to 337.9 million against a weighted population of 327.4 million. The 10.5 million difference has two parts: the boundary ages, counted twice, add 13.1 million, while 2.6 million people (243 respondents with no recorded age) fall in no age group. The two sex groups, by contrast, sum exactly to 327.4 million. 

\subsection{Parameters for the RLSCA}
The parameters used throughout are listed in Table~\ref{tab:rlsca_hyperparameters}. We adopted a pairwise support threshold of $T = 0.0018$ (0.18\% of each group's weighted population: about 600,000 people for the full population and about 60,000--80,000 for each age--sex group) to allow for the discovery of meaningful associations without excluding lower-prevalence clinical profiles.

Because cluster lengths are drawn at random, repeated executions of the
algorithm do not reproduce identical frequency counts. All results
reported here come from a single execution of 100 iterations per stratum.
Across two independent executions under identical settings we found the
same clusters recurring: 15 of the top 20 clusters were identical. The frequency of the same cluster differed
between the two runs by up to ten; for example, the first cluster in
Table~\ref{tab:new_grouping_2019} was returned 20 times in this run and 12 times in the other, while the second was returned 19 times in both. We therefore treat frequency as an ordinal indicator of cluster stability
rather than an exact measurement.

\begin{table}[!htbp]
\centering
\caption{Configuration of Hyperparameters for the RLSCA Algorithm}
\label{tab:rlsca_hyperparameters}
\begin{tabular}{@{}ll@{}}
\toprule
\textbf{Parameter} & \textbf{Configuration Value / Range} \\ \midrule
Data Source & MEPS 2019 (Full Population and Stratified Subgroups) \\
Demographic Strata & Overlapping Intervals: 0--20, 20--40, 40--60, 60+ \\
Gender Stratification & Stratified by Men and Women for each age cohort \\
Pairwise Support Threshold ($T$) & $0.0018$ of each group ($\approx 600{,}000$ people overall) \\
Total Iterations ($M$) & 100 \\
Cluster Cardinality ($R$) & Discrete uniform on $\{3, \dots, 8\}$, drawn at each step \\
Diseases per Iteration ($nd$) & 60, or fewer if no cluster can be formed \\
Final Selection Threshold ($n$) & Top 20 most frequent clusters (input to Algorithm~\ref{alg:nonoverlap}) \\ & (no restriction for Figures~\ref{fig:cluster-women-4060} and~\ref{fig:cluster-men-4060}) \\ \bottomrule
\end{tabular}
\end{table}

\subsection{Selected Clusters for Different Age and Gender Subgroups}

In this section we present one notable cluster for each of eight different age and gender groups (ages $0-20$, $20-40$, $40-60$, and $60+$ further subdivided into male and female). The number of times out of 100 randomized iterations in which the cluster was extracted is also indicated for each cluster.  In addition to the disease names, we also show the name of the broader organ and body system to which the disease belongs. These results confirm that multiple body and organ systems can be involved in multimorbidity. We reiterate that these are notable examples from our cluster database, but not the only clusters in the age and gender subgroup.  

For instance, for women in the group $0-20$, the presented cluster consists of $7$ different diseases that span six different body systems. It is important to note that just because these diseases are together in a cluster does not causally link them together.  However, it does provide epidemiologists and clinicians important clues on possible relationships that are yet to be identified. As an example, it is quite reasonable to hypothesize that a contraceptive or procreative-management diagnosis (176, grouped under pregnancy) is linked to psychological distress, implicating mental health conditions such as anxiety and depression. Also noteworthy is the fact that this large cluster was found 21 times in 100 randomized iterations of the algorithm. 

For women in the $60+$ group, we see non-epithelial cancers (under Neoplasms) grouped with a number of other conditions from other body systems (the cluster was picked 17 times in 100 randomized iterations). Non-melanoma skin cancers are common in this age group. It is therefore worth investigating whether they might be linked to nutritional deficiencies (under Endocrine, Metabolic and Immunity Disorders), skin infections (under Subcutaneous), and/or other bone disorders (under Musculoskeletal), which are part of the large cluster. It is important to note that these other diseases may either be a \emph{precursor} to cancer development or a \emph{consequence} of the treatment of non-epithelial cancers, or completely independent. Whatever the reasons, the cluster provides a clinician with clues on what  conditions and body systems can manifest together, and thereby ensure that different specialties can coordinate the patient's treatment. 

The clusters shown for the remaining age groups in Figures~\ref{fig:clusters0020}–\ref{fig:clusters60} can be read in the same way.  

\begin{figure}[!htbp]
\centering
\begin{tikzpicture}[
  every node/.style={font=\small, anchor=west},
  brace/.style={decorate,decoration={brace,amplitude=8pt,mirror}, thick}
]

\draw [brace] (-6.5,0.9) -- (-6.5,-4.3);
\node[anchor=east] at (-6.7,-1.7) {\textbf{Women 0--20}};
\node[anchor=east] at (-6.7,-2.1) {(21/100)};

\node at (-6.2,0.8) {\textbf{Infectious:}};
\node at (-5.6,0.4) {Mycoses (4)};

\node at (-6.2,0.0) {\textbf{Endocrine/Metabolic:}};
\node at (-5.6,-0.4) {Nutrit defic (52)};

\node at (-6.2,-0.9) {\textbf{Neuro/Sensory:}};
\node at (-5.6,-1.3) {Headache/migraine (84)};

\node at (-6.2,-1.8) {\textbf{Digestive:}};
\node at (-5.6,-2.2) {Ot dx stomch (141)};

\node at (-6.2,-2.7) {\textbf{Pregnancy:}};
\node at (-5.6,-3.1) {Contraceptiv (176)};

\node at (-6.2,-3.6) {\textbf{Mental Health:}};
\node at (-5.6,-4.0) {Anxiety disorders (651)};
\node at (-5.6,-4.4) {Mood disorders (657)};

\draw [brace] (2.4,0.9) -- (2.4,-3.6);
\node[anchor=east] at (2.2,-1.3) {\textbf{Men 0--20}};
\node[anchor=east] at (2.2,-1.7) {(12/100)};

\node at (2.7,0.8) {\textbf{Neuro/Sensory:}};
\node at (3.3,0.4) {Blindness (89)};

\node at (2.7,0.0) {\textbf{Respiratory:}};
\node at (3.3,-0.4) {Influenza (123)};

\node at (2.7,-0.9) {\textbf{Digestive:}};
\node at (3.3,-1.3) {Other GI dx (155)};

\node at (2.7,-1.8) {\textbf{Skin:}};
\node at (3.3,-2.2) {Oth skin dx (200)};

\node at (2.7,-2.7) {\textbf{Injury:}};
\node at (3.3,-3.1) {Fracture of arm (229)};

\end{tikzpicture}
\caption{Multimorbidity clusters for Men and Women aged 0–20, grouped by Level 1 CCC categories (codes in parentheses).}
\label{fig:clusters0020}
\end{figure}

\begin{figure}[!htbp]
\centering
\begin{tikzpicture}[
  every node/.style={font=\small, anchor=west},
  brace/.style={decorate,decoration={brace,amplitude=8pt,mirror}, thick}
]

\draw [brace] (-6.5,0.9) -- (-6.5,-4.3);
\node[anchor=east] at (-6.7,-1.7) {\textbf{Women 20--40}};
\node[anchor=east] at (-6.7,-2.1) {(27/100)};

\node at (-6.2,0.8) {\textbf{Endocrine/Metabolic:}};
\node at (-5.6,0.4) {Diabetes (49)};
\node at (-5.6,0.0) {High Cholesterol (53)};

\node at (-6.2,-0.5) {\textbf{Neuro/Sensory:}};
\node at (-5.6,-0.9) {Oth nerv dx (95)};

\node at (-6.2,-1.4) {\textbf{Circulatory:}};
\node at (-5.6,-1.8) {Hypertension (98)};

\node at (-6.2,-2.3) {\textbf{Digestive:}};
\node at (-5.6,-2.7) {Esophgeal dx (138)};

\node at (-6.2,-3.2) {\textbf{Musculoskeletal:}};
\node at (-5.6,-3.6) {Osteoarthritis (203)};
\node at (-5.6,-4.0) {Ot conn tiss (211)};

\draw [brace] (2.4,0.9) -- (2.4,-5.3);
\node[anchor=east] at (2.2,-2.2) {\textbf{Men 20--40}};
\node[anchor=east] at (2.2,-2.6) {(14/100)};

\node at (2.7,0.8) {\textbf{Endocrine/Metabolic:}};
\node at (3.3,0.4) {High Cholesterol (53)};
\node at (3.3,0.0) {Diabetes (49)};

\node at (2.7,-0.5) {\textbf{Neuro/Sensory:}};
\node at (3.3,-0.9) {Oth nerv dx (95)};

\node at (2.7,-1.4) {\textbf{Circulatory:}};
\node at (3.3,-1.8) {Hypertension (98)};

\node at (2.7,-2.3) {\textbf{Respiratory:}};
\node at (3.3,-2.7) {Oth uppr resp (134)};

\node at (2.7,-3.2) {\textbf{Digestive:}};
\node at (3.3,-3.6) {Esophgeal dx (138)};

\node at (2.7,-4.1) {\textbf{Mental Health:}};
\node at (3.3,-4.5) {Anxiety disorders (651)};
\node at (3.3,-4.9) {Mood disorders (657)};

\end{tikzpicture}
\caption{Multimorbidity clusters for Men and Women aged 20–40, grouped by Level 1 CCC categories (codes in parentheses).}
\label{fig:clusters2040}
\end{figure}

\begin{figure}[!htbp]
\centering
\begin{tikzpicture}[
  every node/.style={font=\small, anchor=west},
  brace/.style={decorate,decoration={brace,amplitude=8pt,mirror}, thick}
]

\draw [brace] (-6.5,0.9) -- (-6.5,-3.5);
\node[anchor=east] at (-6.7,-1.2) {\textbf{Women 40--60}};
\node[anchor=east] at (-6.7,-1.6) {(9/100)};

\node at (-6.2,0.8) {\textbf{Circulatory:}};
\node at (-5.6,0.4) {Acute CVD (109)};
\node at (-5.6,0.0) {Ot circul dx (117)};

\node at (-6.2,-0.4) {\textbf{Respiratory:}};
\node at (-5.6,-0.8) {Ot up rsp in (126)};

\node at (-6.2,-1.2) {\textbf{Genitourinary:}};
\node at (-5.6,-1.6) {Menopausl dx (173)};

\node at (-6.2,-2.0) {\textbf{Musculoskeletal:}};
\node at (-5.6,-2.4) {Back problem (205)};

\node at (-6.2,-2.8) {\textbf{Injury and Poisoning:}};
\node at (-5.6,-3.2) {Other injury (244)};

\draw [brace] (2.4,0.9) -- (2.4,-3.8);
\node[anchor=east] at (2.2,-1.3) {\textbf{Men 40--60}};
\node[anchor=east] at (2.2,-1.7) {(12/100)};

\node at (2.7,0.8) {\textbf{Neuro/Sensory:}};
\node at (3.3,0.4) {Headache/mig (84)};

\node at (2.7,0.0) {\textbf{Respiratory:}};
\node at (3.3,-0.4) {Ot uppr resp (134)};

\node at (2.7,-0.8) {\textbf{Digestive:}};
\node at (3.3,-1.2) {Ot dx stomch (141)};
\node at (3.3,-1.6) {Other GI dx (155)};

\node at (2.7,-2.0) {\textbf{Musculoskeletal:}};
\node at (3.3,-2.4) {Ot joint dx (204)};

\node at (2.7,-2.8) {\textbf{Injury and Poisoning:}};
\node at (3.3,-3.2) {Joint injury (225)};
\node at (3.3,-3.6) {Other injury (244)};

\end{tikzpicture}
\caption{Multimorbidity clusters for Men and Women aged 40–60, grouped by Level 1 CCC categories (codes in parentheses).}
\label{fig:clusters4060}
\end{figure}

\begin{figure}[!htbp]
\centering
\begin{tikzpicture}[
  every node/.style={font=\small, anchor=west},
  brace/.style={decorate,decoration={brace,amplitude=8pt,mirror}, thick}
]

\draw [brace] (-6.5,0.8) -- (-6.5,-4.0);
\node[anchor=east] at (-6.7,-1.6) {\textbf{Women 60+}};
\node[anchor=east] at (-6.7,-2.0) {(17/100)};

\node at (-6.2,0.8) {\textbf{Neoplasms:}};
\node at (-5.6,0.4) {Non-epith skin ca (23)};

\node at (-6.2,0.0) {\textbf{Endocrine/Metabolic:}};
\node at (-5.6,-0.4) {Nutritional deficiencies (52)};

\node at (-6.2,-0.8) {\textbf{Neuro/Sensory:}};
\node at (-5.6,-1.2) {Headache/migraine (84)};

\node at (-6.2,-1.6) {\textbf{Skin/Subcutaneous:}};
\node at (-5.6,-2.0) {Skin infection (197)};
\node at (-5.6,-2.4) {Ot infl skin (198)};

\node at (-6.2,-2.8) {\textbf{Musculoskeletal:}};
\node at (-5.6,-3.2) {Bone disorders (212)};

\node at (-6.2,-3.6) {\textbf{Injury and Poisoning:}};
\node at (-5.6,-4.0) {Other injury (244)};

\draw [brace] (2.4,0.8) -- (2.4,-4.0);
\node[anchor=east] at (2.2,-1.6) {\textbf{Men 60+}};
\node[anchor=east] at (2.2,-2.0) {(18/100)};

\node at (2.7,0.8) {\textbf{Neuro/Sensory:}};
\node at (3.3,0.4) {Eye infectn (90)};
\node at (3.3,0.0) {Oth nerv dx (95)};

\node at (2.7,-0.4) {\textbf{Respiratory:}};
\node at (3.3,-0.8) {Pneumonia (122)};
\node at (3.3,-1.2) {COPD (127)};
\node at (3.3,-1.6) {Other lower resp (133)};

\node at (2.7,-2.0) {\textbf{Digestive:}};
\node at (3.3,-2.4) {Other GI disorders (155)};

\node at (2.7,-2.8) {\textbf{Genitourinary:}};
\node at (3.3,-3.2) {Renal failure (157)};

\node at (2.7,-3.6) {\textbf{Musculoskeletal:}};
\node at (3.3,-4.0) {Ot bone dx (212)};

\end{tikzpicture}
\caption{Multimorbidity clusters for Men and Women aged 60+, grouped by Level 1 CCC categories (codes in parentheses).}
\label{fig:clusters60}
\end{figure}

\FloatBarrier
\subsection{Final Clustering Solutions and Visualizations}

We first apply Algorithm~\ref{alg:nonoverlap} to the full 2019 population. Using the cluster
database generated by Algorithm~\ref{alg:RLSCA} at a threshold of $T=0.0018$, the final set of
non-overlapping clusters is shown in Table~\ref{tab:new_grouping_2019}. Note that these clusters
differ from the high-frequency examples of Section 3.3, which used the higher threshold $T=0.018$. They also differ because clusters are sorted by
length first, and once a cluster is chosen, any shorter cluster sharing even one disease with it
can no longer be picked.
Cluster 1 consists
of well known diseases that further provide clinical validity to our approach: the well known
trio of hypertension, high cholesterol and diabetes is linked to ``downstream'' cardiac and
cerebrovascular conditions such as buildup of plaque in the coronary arteries (atherosclerosis),
heart attack (myocardial infarction), and mini-stroke (transient ischemic attack). Cluster 2, on the
other hand, is less known, and consists of eight diseases that span five different body systems:
eye disorders such as cataract are grouped with cardiac problems, chronic obstructive pulmonary
(lung) disease, esophageal and digestive disorders, and osteoarthritis. Since the cluster is
chosen frequently in our algorithm, the relationships between diseases in it are worthy of
further clinical investigation.
\begin{table}[!htbp]
\centering
\caption{Disease clusters for the full population (2019) with disease names, diversity
values, and frequencies. Numbers in parentheses are CCC codes; Table~\ref{bodysystems} lists them by body system.}
\label{tab:new_grouping_2019}
\begin{tabular}{|c|>{\raggedright\arraybackslash}p{10.5cm}|c|c|}
\hline
\textbf{Cluster} & \textbf{Diseases (Name + Code)} & \textbf{Div.} & \textbf{Freq.} \\
\hline
1 & DiabMel no c (49), Hyperlipidem (53), Hypertension (98), Acute MI (100), Coronary atherosclerosis and other heart
disease (101), Oth heart dx (104), chf;nonhp (108), TIA (112) & 2 & 20 \\
\hline
2 & Cataract (86), Other eye dx (91), Oth nerv dx (95), Cardiac dysrhythmias (106), COPD (127), Esophgeal dx
(138), Other GI dx (155), Osteoarthritis (203) & 5 & 19 \\
\hline
3 & Asthma (128), Oth low resp (133), Ot uppr resp (134), Rheum arth (202), Other injury (244) & 3 & 5 \\
\hline
4 & Viral infect (7), Influenza (123), Ot up rsp in (126) & 2 & 10 \\
\hline
5 & Ot joint dx (204), Back problem (205), Ot bone dx (212) & 1 & 9 \\
\hline
6 & Anxiety disorders (651), Attention-deficit/conduct/disruptive behavior disorders (652), Mood disorders (657) & 1 & 5 \\
\hline
\end{tabular}
\end{table}

\begin{figure}[htbp]
  \centering
  \includegraphics{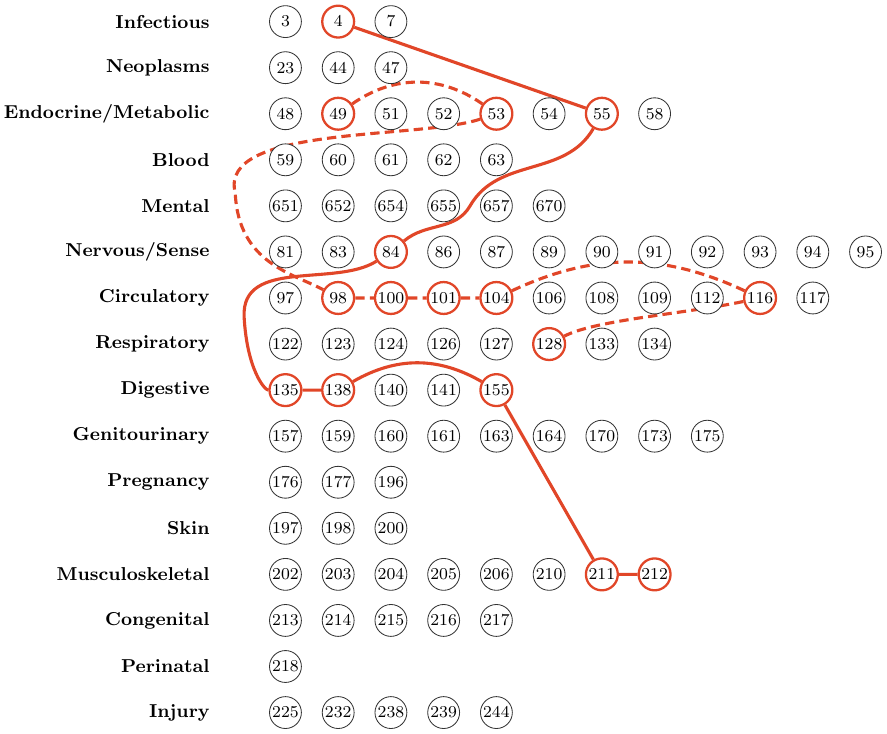} 
  \caption{Final disease cluster examples — Women, 40–60. Each line joins the conditions of one cluster (thick circles); Table~\ref{bodysystems} gives the disease name for each code. Drawn with TikZ code written with the help of Anthropic’s Claude Opus 5.5 and checked by the authors.}
  \label{fig:cluster-women-4060}
\end{figure}

Having established the full-population solution, we now turn to one demographic stratum,
women aged $40-60$. The previous section presented exploratory results involving large and
diverse clusters that showed up consistently in RLSCA across different age and gender
demographic groups. These clusters, while they occur frequently, contain overlapping
diseases. In this section, we illustrate how the RLSCA creates non-overlapping clusters for
this subgroup by applying Algorithm~\ref{alg:nonoverlap} to the comprehensive cluster
database generated for it. Understanding disease clusters in ``middle-age'' provides
clinicians and epidemiologists with tools to prevent and treat chronic conditions early.
For Figure~\ref{fig:cluster-women-4060} we depart from the $n=20$ selection used elsewhere and apply
Algorithm~\ref{alg:nonoverlap} to the full cluster database, without restricting the
candidate set. Sorting by length and breaking ties by frequency, we take the first two
clusters that share no diseases with one another. Restricting the candidates to the twenty
most frequent clusters favors short, common groupings; using the full database instead lets
the longest and most systemically diverse clusters through, which is what the figure is
intended to show. The result is a pair of eight-disease clusters. The corresponding
solution for men aged $40-60$ is given in the Appendix (Figure~\ref{fig:cluster-men-4060}).

Figure~\ref{fig:cluster-women-4060} shows the final clusters for women aged $40-60$. In
this visualization, the diseases are organized in rows, with each row corresponding to the
16 different body systems. Each row consists of numbers in circles which refer to the
diseases within that body system, and each cluster is drawn as a line connecting its
diseases. The dashed cluster is the cardio-metabolic group -- diabetes (49), lipid
disorders (53) and hypertension (98) together with myocardial infarction (100), coronary
atherosclerosis (101), other heart disease (104) and arterial embolism (116) -- and it also
includes asthma (128), a respiratory condition. It is the most frequently rediscovered
eight-disease cluster in this stratum. The solid cluster spans five body systems and links
mycoses (fungal infections, 4), a fluid and electrolyte disorder (55),
headache/migraines (84), intestinal infection (135), esophageal (138) and other
gastrointestinal (155) disorders, and two musculoskeletal conditions, connective tissue
disease (211) and bone disorders (212). Cross-system groupings of this kind are what the
database is intended to surface as candidates for further clinical investigation.

\section{ Conclusions and Future Research}

In summary, we have demonstrated a new methodology, the RLSCA, to identify key disease association patterns in multimorbidity. Using age and sex subgroups in the nationally representative MEPS dataset, we have shown how the method can reveal clusters that span multiple body systems. We have also compared the RLSCA with existing clustering algorithms, illustrating its strengths. 

There are several directions for future research. Currently our algorithm analyzes disease combinations in the US population at a single point of time. A more powerful enhancement of the algorithm is considering the \textit{longitudinal progression}  of diseases within an individual over many decades. The order and age at which diseases develop in an individual provides vital clues for disease prevention. A second enhancement would be the inclusion of biomarkers, medications, procedures and treatments as part of the feature space. This gives a far more holistic view than age, gender, and the presence/absence of diseases that we currently illustrate in this paper. However, this also requires significant adaptation of the RLSCA to include mixed variable types, since many biomarkers (such as blood pressure, heart rate, blood sugar) fall in continuous ranges.  

While these are important future research directions, we note that our algorithm and its results can already be clinically useful. Because of the sheer heterogeneity of chronic conditions, physicians routinely encounter patients whose particular combination of chronic conditions falls outside existing clinical guidelines or their own prior experience. Although medical large language models (LLMs) can summarize the published literature, clinicians currently lack tools that identify clinically meaningful disease patterns directly from patient-level data and are tailored to an individual patient's multimorbidity profile. Results from our algorithm trained on the MEPS dataset can be integrated into a clinical decision support tool and combined with a trusted medical LLM such as OpenEvidence, which will allow physicians to query the validity of the clusters produced by our algorithm. This is particularly useful for primary care physicians who practice individually or in small groups, or who do not have access to larger electronic health record databases. For larger health systems, the patient-level chronic conditions in the EHR can be used as an input to our clustering algorithm. EHRs of integrated health systems such as the Veterans Affairs (VA), or those built on vendor platforms such as Epic and Cerner, are likely to yield far more reliable clusters than MEPS, due to the sheer volume of data. 

\section*{Data availability statement}
The Medical Expenditure Panel Survey data used in this study are publicly available from the
Agency for Healthcare Research and Quality at https://meps.ahrq.gov. Code implementing the RLSCA,
the hierarchical clustering benchmark, and the cluster databases generated in this study are
available from the authors on request and will be deposited in a public repository (Zenodo, with a DOI) when the paper is published.

\section*{Declaration of generative AI use}
The authors used Anthropic’s Claude Opus 5 and Claude Opus 5.5 as writing and checking aids during the
preparation of this manuscript: to proofread and copy-edit text written by the authors and propose wording for some passages, to suggest which material to shorten or move to the Appendix to meet the journal's page limit, to shorten the authors' abstract to the 200-word limit, to help write the plotting code for Figures 2, 7 and 8, to
cross-check reported cluster statistics against the underlying data files and compute some of the descriptive statistics reported in the text, and to check
citations against their sources. The algorithm, the model formulation, the analysis and all
scientific claims are the authors’ own. The authors reviewed and verified all content and take
full responsibility for it, including the accuracy of the references. The authors reviewed the terms of use of the tool and confirmed that its use was appropriate for this manuscript.

\section*{Funding}
This research was funded by the U.S. National Science Foundation grant NSF CMMI 2212237. The views expressed in this paper are those of the authors and not of the National Science Foundation.

\section*{Declaration of interest}
The authors report no conflict of interest.

\spacingset{1}\printbibliography

@article{zhang2019multimorbidity,
	chapter = {Epidemiology},
	title = {Multimorbidity of cardiometabolic diseases: prevalence and risk for mortality from one million {Chinese} adults in a longitudinal cohort study},
	volume = {9},
	copyright = {© Author(s) (or their employer(s)) 2019. Re-use permitted under CC BY-NC. No commercial re-use. See rights and permissions. Published by BMJ.. This is an open access article distributed in accordance with the Creative Commons Attribution Non Commercial (CC BY-NC 4.0) license, which permits others to distribute, remix, adapt, build upon this work non-commercially, and license their derivative works on different terms, provided the original work is properly cited, appropriate credit is given, any changes made indicated, and the use is non-commercial. See: http://creativecommons.org/licenses/by-nc/4.0/.},
	issn = {2044-6055, 2044-6055},
	shorttitle = {Multimorbidity of cardiometabolic diseases},
	url = {https://bmjopen.bmj.com/content/9/3/e024476},
	doi = {10.1136/bmjopen-2018-024476},
	language = {en},
	number = {3},
	urldate = {2026-09-23},
	journal = {BMJ Open},
	publisher = {British Medical Journal Publishing Group},
	author = {Zhang, Dudan and Tang, Xun and Shen, Peng and Si, Yaqin and Liu, Xiaofei and Xu, Zhe and Wu, Jinguo and Zhang, Jingyi and Lu, Ping and Lin, Hongbo and Gao, Pei},
	month = mar,
	year = {2019},
	pages = {e024476},
}

@article{valles2019neuroactive,
	title = {The neuroactive potential of the human gut microbiota in quality of life and depression},
	volume = {4},
	copyright = {2019 The Author(s), under exclusive licence to Springer Nature Limited},
	issn = {2058-5276},
	url = {https://www.nature.com/articles/s41564-018-0337-x},
	doi = {10.1038/s41564-018-0337-x},
	language = {en},
	number = {4},
	urldate = {2026-09-23},
	journal = {Nature Microbiology},
	publisher = {Nature Publishing Group},
	author = {Valles-Colomer, Mireia and Falony, Gwen and Darzi, Youssef and Tigchelaar, Ettje F. and Wang, Jun and Tito, Raul Y. and Schiweck, Carmen and Kurilshikov, Alexander and Joossens, Marie and Wijmenga, Cisca and Claes, Stephan and Van Oudenhove, Lukas and Zhernakova, Alexandra and Vieira-Silva, Sara and Raes, Jeroen},
	month = apr,
	year = {2019},
	pages = {623--632},
}

@article{vaghfi2025identifying,
	title = {Identifying most lethal cliques in disease comorbidity graphs},
	volume = {15},
	issn = {2472-5579},
	url = {https://www.tandfonline.com/doi/abs/10.1080/24725579.2025.2473716},
	doi = {10.1080/24725579.2025.2473716},
	number = {2},
	urldate = {2026-09-23},
	journal = {IISE Transactions on Healthcare Systems Engineering},
	publisher = {Taylor \& Francis},
	author = {Vaghfi Mohebbi, Parisa and Lu, Yajun and Miao, Zhuqi and Balasundaram, Balabhaskar and Kalgotra, Pankush and Sharda, Ramesh},
	month = apr,
	year = {2025},
	pages = {183--200},
}

@article{thirunavukarasu2023large,
	title = {Large language models in medicine},
	volume = {29},
	copyright = {2023 Springer Nature America, Inc.},
	issn = {1546-170X},
	url = {https://www.nature.com/articles/s41591-023-02448-8},
	doi = {10.1038/s41591-023-02448-8},
	language = {en},
	number = {8},
	urldate = {2026-09-23},
	journal = {Nature Medicine},
	publisher = {Nature Publishing Group},
	author = {Thirunavukarasu, Arun James and Ting, Darren Shu Jeng and Elangovan, Kabilan and Gutierrez, Laura and Tan, Ting Fang and Ting, Daniel Shu Wei},
	month = aug,
	year = {2023},
	pages = {1930--1940},
}

@article{sorace2011complexity,
	title = {The complexity of disease combinations in the {Medicare} population},
	volume = {14},
	issn = {1942-7905},
	doi = {10.1089/pop.2010.0044},
	language = {eng},
	number = {4},
	journal = {Population Health Management},
	author = {Sorace, James and Wong, Hui-Hsing and Worrall, Chris and Kelman, Jeffrey and Saneinejad, Shahin and MaCurdy, Thomas},
	month = aug,
	year = {2011},
	pages = {161--166},
}

@techreport{sorace2013understanding,
title = {Understanding the High Prevalence of Low-Prevalence Chronic Disease Combinations: {Databases} and {
Methods} for {Research}},
author = {Rezaee, Michael E. and LeRoy, Lisa and White, Alan and Oppenheim, Emma and Carlson, Ken and Wasserman,
Melanie},
institution = {{Abt Associates Inc., prepared for the Office of the Assistant Secretary for Planning and
Evaluation, U.S. Department of Health and Human Services}},
type = {Final White Paper},
year = {2013},
url = {https://aspe.hhs.gov/reports/understanding-high-prevalence-low-prevalence-chronic-disease-combinations-databases-methods-research-0}
}

@article{soley2021identifying,
	title = {Identifying multimorbidity clusters with the highest primary care use: 15 years of evidence from a multi-ethnic metropolitan population},
	volume = {72},
	copyright = {http://creativecommons.org/licenses/by/4.0/},
	issn = {0960-1643, 1478-5242},
	shorttitle = {Identifying multimorbidity clusters with the highest primary care use},
	url = {https://bjgp.org/lookup/doi/10.3399/BJGP.2021.0325},
	doi = {10.3399/BJGP.2021.0325},
	language = {en},
	number = {716},
	urldate = {2026-09-23},
	journal = {British Journal of General Practice},
	author = {Soley-Bori, Marina and Bisquera, Alessandra and Ashworth, Mark and Wang, Yanzhong and Durbaba, Stevo and Dodhia, Hiten and Fox-Rushby, Julia},
	month = mar,
	year = {2022},
	pages = {e190--e198},
}

@article{skou2022multimorbidity,
	title = {Multimorbidity},
	volume = {8},
	copyright = {2022 Springer Nature Limited},
	issn = {2056-676X},
	url = {https://www.nature.com/articles/s41572-022-00376-4},
	doi = {10.1038/s41572-022-00376-4},
	language = {en},
	number = {1},
	urldate = {2026-09-23},
	journal = {Nature Reviews Disease Primers},
	publisher = {Nature Publishing Group},
	author = {Skou, Søren T. and Mair, Frances S. and Fortin, Martin and Guthrie, Bruce and Nunes, Bruno P. and Miranda, J. Jaime and Boyd, Cynthia M. and Pati, Sanghamitra and Mtenga, Sally and Smith, Susan M.},
	month = jul,
	year = {2022},
	pages = {48},
}

@article{singhal2023large,
	title = {Large language models encode clinical knowledge},
	volume = {620},
	copyright = {2023 The Author(s)},
	issn = {1476-4687},
	url = {https://www.nature.com/articles/s41586-023-06291-2},
	doi = {10.1038/s41586-023-06291-2},
	language = {en},
	number = {7972},
	urldate = {2026-09-23},
	journal = {Nature},
	publisher = {Nature Publishing Group},
	author = {Singhal, Karan and Azizi, Shekoofeh and Tu, Tao and Mahdavi, S. Sara and Wei, Jason and Chung, Hyung Won and Scales, Nathan and Tanwani, Ajay and Cole-Lewis, Heather and Pfohl, Stephen and Payne, Perry and Seneviratne, Martin and Gamble, Paul and Kelly, Chris and Babiker, Abubakr and Schärli, Nathanael and Chowdhery, Aakanksha and Mansfield, Philip and Demner-Fushman, Dina and Agüera y Arcas, Blaise and Webster, Dale and Corrado, Greg S. and Matias, Yossi and Chou, Katherine and Gottweis, Juraj and Tomasev, Nenad and Liu, Yun and Rajkomar, Alvin and Barral, Joelle and Semturs, Christopher and Karthikesalingam, Alan and Natarajan, Vivek},
	month = aug,
	year = {2023},
	pages = {172--180},
}

@article{kruskal1964multidimensional,
	title = {Multidimensional {Scaling} by {Optimizing} {Goodness} of {Fit} to a {Nonmetric} {Hypothesis}},
	volume = {29},
	issn = {0033-3123, 1860-0980},
	url = {https://www.cambridge.org/core/journals/psychometrika/article/abs/multidimensional-scaling-by-optimizing-goodness-of-fit-to-a-nonmetric-hypothesis/60D38EAAC9DB3FD765AEDBECEB4F4333},
	doi = {10.1007/BF02289565},
	language = {en},
	number = {1},
	urldate = {2026-09-23},
	journal = {Psychometrika},
	author = {Kruskal, J. B.},
	month = mar,
	year = {1964},
	pages = {1--27},
}

@article{robertson2022identifying,
	title = {Identifying multimorbidity clusters in an unselected population of hospitalised patients},
	volume = {12},
	copyright = {2022 The Author(s)},
	issn = {2045-2322},
	url = {https://www.nature.com/articles/s41598-022-08690-3},
	doi = {10.1038/s41598-022-08690-3},
	language = {en},
	number = {1},
	urldate = {2026-09-23},
	journal = {Scientific Reports},
	publisher = {Nature Publishing Group},
	author = {Robertson, Lynn and Vieira, Rute and Butler, Jessica and Johnston, Marjorie and Sawhney, Simon and Black, Corri},
	month = mar,
	year = {2022},
	pages = {5134},
}

@article{reaven1988role,
	title = {Role of {Insulin} {Resistance} in {Human} {Disease}},
	volume = {37},
	issn = {0012-1797},
	url = {https://doi.org/10.2337/diab.37.12.1595},
	doi = {10.2337/diab.37.12.1595},
	number = {12},
	urldate = {2026-09-23},
	journal = {Diabetes},
	author = {Reaven, Gerald M},
	month = dec,
	year = {1988},
	pages = {1595--1607},
}

@article{prados2014multimorbidity,
	title = {Multimorbidity patterns: a systematic review},
	volume = {67},
	issn = {0895-4356},
	shorttitle = {Multimorbidity patterns},
	url = {https://www.sciencedirect.com/science/article/pii/S0895435613004368},
	doi = {10.1016/j.jclinepi.2013.09.021},
	number = {3},
	urldate = {2026-09-23},
	journal = {Journal of Clinical Epidemiology},
	author = {Prados-Torres, Alexandra and Calderón-Larrañaga, Amaia and Hancco-Saavedra, Jorge and Poblador-Plou, Beatriz and van den Akker, Marjan},
	month = mar,
	year = {2014},
	pages = {254--266},
}

@article{poblador2014similar,
	title = {Similar {Multimorbidity} {Patterns} in {Primary} {Care} {Patients} from {Two} {European} {Regions}: {Results} of a {Factor} {Analysis}},
	volume = {9},
	issn = {1932-6203},
	shorttitle = {Similar {Multimorbidity} {Patterns} in {Primary} {Care} {Patients} from {Two} {European} {Regions}},
	url = {https://journals.plos.org/plosone/article?id=10.1371/journal.pone.0100375},
	doi = {10.1371/journal.pone.0100375},
	language = {en},
	number = {6},
	urldate = {2026-09-23},
	journal = {PLOS ONE},
	publisher = {Public Library of Science},
	author = {Poblador-Plou, Beatriz and Akker, Marjan van den and Vos, Rein and Calderón-Larrañaga, Amaia and Metsemakers, Job and Prados-Torres, Alexandra},
	month = jun,
	year = {2014},
	pages = {e100375},
}

@article{peng2023association,
	title = {Association between thyroid dysfunction, metabolic disturbances, and clinical symptoms in first-episode, untreated {Chinese} patients with major depressive disorder: {Undirected} and {Bayesian} network analyses},
	volume = {14},
	issn = {1664-2392},
	shorttitle = {Association between thyroid dysfunction, metabolic disturbances, and clinical symptoms in first-episode, untreated {Chinese} patients with major depressive disorder},
	url = {https://www.frontiersin.org/journals/endocrinology/articles/10.3389/fendo.2023.1138233/full},
	doi = {10.3389/fendo.2023.1138233}, eid = {1138233},
	language = {English},
	urldate = {2026-09-23},
	journal = {Frontiers in Endocrinology},
	publisher = {Frontiers},
	author = {Peng, Pu and Wang, Qianjin and Lang, Xiao E. and Liu, Tieqiao and Zhang, Xiang-Yang},
	month = feb,
	year = {2023},
}

@article{oken1999issue,
	title = {At {Issue}: {Schizophrenia} and {Rheumatoid} {Arthritis}: {The} {Negative} {Association} {Revisited}},
	volume = {25},
	issn = {0586-7614},
	shorttitle = {At {Issue}},
	url = {https://doi.org/10.1093/oxfordjournals.schbul.a033407},
	doi = {10.1093/oxfordjournals.schbul.a033407},
	number = {4},
	urldate = {2026-09-23},
	journal = {Schizophrenia Bulletin},
	author = {Oken, Robert J. and Schulzer, Michael},
	month = jan,
	year = {1999},
	pages = {625--638},
}

@article{ng2018patterns,
	title = {Patterns of multimorbid health conditions: a systematic review of analytical methods and comparison analysis},
	volume = {47},
	issn = {0300-5771},
	shorttitle = {Patterns of multimorbid health conditions},
	url = {https://doi.org/10.1093/ije/dyy134},
	doi = {10.1093/ije/dyy134},
	number = {5},
	urldate = {2026-09-23},
	journal = {International Journal of Epidemiology},
	author = {Ng, Shu Kay and Tawiah, Richard and Sawyer, Michael and Scuffham, Paul},
	month = oct,
	year = {2018},
	pages = {1687--1704},
}

@article{li2016metabolic,
	title = {Metabolic syndrome and components exacerbate osteoarthritis symptoms of pain, depression and reduced knee function},
	volume = {4},
	issn = {2305-5839},
	url = {https://pmc.ncbi.nlm.nih.gov/articles/PMC4842398/},
	doi = {10.21037/atm.2016.03.48},
	number = {7},
	urldate = {2026-09-23},
	journal = {Annals of Translational Medicine},
	author = {Li, Hongxing and George, Daniel M. and Jaarsma, Ruurd L. and Mao, Xinzhan},
	month = apr,
	year = {2016},
	pages = {133},
}

@article{lenzi2016burden,
	chapter = {Epidemiology},
	title = {Burden of multimorbidity in relation to age, gender and immigrant status: a cross-sectional study based on administrative data},
	volume = {6},
	copyright = {Published by the BMJ Publishing Group Limited. For permission to use (where not already granted under a licence) please go to http://www.bmj.com/company/products-services/rights-and-licensing/. This is an Open Access article distributed in accordance with the terms of the Creative Commons Attribution (CC BY 4.0) license, which permits others to distribute, remix, adapt and build upon this work, for commercial use, provided the original work is properly cited. See: http://creativecommons.org/licenses/by/4.0/},
	issn = {2044-6055, 2044-6055},
	shorttitle = {Burden of multimorbidity in relation to age, gender and immigrant status},
	url = {https://bmjopen.bmj.com/content/6/12/e012812},
	doi = {10.1136/bmjopen-2016-012812},
	language = {en},
	number = {12},
	urldate = {2026-09-23},
	journal = {BMJ Open},
	publisher = {British Medical Journal Publishing Group},
	author = {Lenzi, Jacopo and Avaldi, Vera Maria and Rucci, Paola and Pieri, Giulia and Fantini, Maria Pia},
	month = dec,
	year = {2016},
	pages = {e012812},
}

@article{kivimaki2017overweight,
	title = {Overweight, obesity, and risk of cardiometabolic multimorbidity: pooled analysis of individual-level data for 120 813 adults from 16 cohort studies from the {USA} and {Europe}},
	volume = {2},
	issn = {2468-2667},
	shorttitle = {Overweight, obesity, and risk of cardiometabolic multimorbidity},
	url = {https://www.thelancet.com/journals/lanpub/article/PIIS2468-2667(17)30074-9/fulltext},
	doi = {10.1016/S2468-2667(17)30074-9},
	language = {English},
	number = {6},
	urldate = {2026-09-23},
	journal = {The Lancet Public Health},
	publisher = {Elsevier},
	author = {Kivimäki, Mika and Kuosma, Eeva and Ferrie, Jane E. and Luukkonen, Ritva and Nyberg, Solja T. and Alfredsson, Lars and Batty, G. David and Brunner, Eric J. and Fransson, Eleonor and Goldberg, Marcel and Knutsson, Anders and Koskenvuo, Markku and Nordin, Maria and Oksanen, Tuula and Pentti, Jaana and Rugulies, Reiner and Shipley, Martin J. and Singh-Manoux, Archana and Steptoe, Andrew and Suominen, Sakari B. and Theorell, Töres and Vahtera, Jussi and Virtanen, Marianna and Westerholm, Peter and Westerlund, Hugo and Zins, Marie and Hamer, Mark and Bell, Joshua A. and Tabak, Adam G. and Jokela, Markus},
	month = jun,
	year = {2017},
	pages = {e277--e285},
}

@article{kessler2005prevalence,
	title = {Prevalence, {Severity}, and {Comorbidity} of 12-{Month} {DSM}-{IV} {Disorders} in the {National} {Comorbidity} {Survey} {Replication}},
	volume = {62},
	issn = {0003-990X},
	url = {https://doi.org/10.1001/archpsyc.62.6.617},
	doi = {10.1001/archpsyc.62.6.617},
	number = {6},
	urldate = {2026-09-23},
	journal = {Archives of General Psychiatry},
	author = {Kessler, Ronald C. and Chiu, Wai Tat and Demler, Olga and Walters, Ellen E.},
	month = jun,
	year = {2005},
	pages = {617--627},
}

@article{kannel1979diabetes,
	title = {Diabetes and {Cardiovascular} {Disease}: {The} {Framingham} {Study}},
	volume = {241},
	issn = {0098-7484},
	shorttitle = {Diabetes and {Cardiovascular} {Disease}},
	url = {https://doi.org/10.1001/jama.1979.03290450033020},
	doi = {10.1001/jama.1979.03290450033020},
	number = {19},
	urldate = {2026-09-23},
	journal = {JAMA},
	author = {Kannel, William B. and McGee, Daniel L.},
	month = may,
	year = {1979},
	pages = {2035--2038},
}

@article{hurt2025use,
	title = {The {Use} of an {Artificial} {Intelligence} {Platform} {OpenEvidence} to {Augment} {Clinical} {Decision}-{Making} for {Primary} {Care} {Physicians}},
	volume = {16},
	issn = {2150-1327},
	doi = {10.1177/21501319251332215},
	language = {eng},
	journal = {Journal of Primary Care \& Community Health},
	author = {Hurt, Ryan T. and Stephenson, Christopher R. and Gilman, Elizabeth A. and Aakre, Christopher A. and Croghan, Ivana T. and Mundi, Manpreet S. and Ghosh, Karthik and Edakkanambeth Varayil, Jithinraj},
	year = {2025},
	pages = {21501319251332215},
}

@article{han2000mining,
	title = {Mining frequent patterns without candidate generation},
	volume = {29},
	issn = {0163-5808},
	url = {https://dl.acm.org/doi/10.1145/335191.335372},
	doi = {10.1145/335191.335372},
	number = {2},
	urldate = {2026-09-23},
	journal = {ACM SIGMOD Record},
	author = {Han, Jiawei and Pei, Jian and Yin, Yiwen},
	month = may,
	year = {2000},
	pages = {1--12},
}

@article{goldstein2009comorbidity,
	title = {Comorbidity {Between} {Psychiatric} and {General} {Medical} {Disorders} in {Homeless} {Veterans}},
	volume = {80},
	issn = {1573-6709},
	url = {https://doi.org/10.1007/s11126-009-9106-6},
	doi = {10.1007/s11126-009-9106-6},
	language = {en},
	number = {4},
	urldate = {2026-09-23},
	journal = {Psychiatric Quarterly},
	author = {Goldstein, Gerald and Luther, James F. and Haas, Gretchen L. and Gordon, Adam J. and Appelt, Cathleen},
	month = dec,
	year = {2009},
	pages = {199--212},
}

@article{garin2016global,
	title = {Global {Multimorbidity} {Patterns}: {A} {Cross}-{Sectional}, {Population}-{Based}, {Multi}-{Country} {Study}},
	volume = {71},
	issn = {1079-5006},
	shorttitle = {Global {Multimorbidity} {Patterns}},
	url = {https://doi.org/10.1093/gerona/glv128},
	doi = {10.1093/gerona/glv128},
	number = {2},
	urldate = {2026-09-23},
	journal = {The Journals of Gerontology: Series A},
	author = {Garin, Noe and Koyanagi, Ai and Chatterji, Somnath and Tyrovolas, Stefanos and Olaya, Beatriz and Leonardi, Matilde and Lara, Elvira and Koskinen, Seppo and Tobiasz-Adamczyk, Beata and Ayuso-Mateos, Jose Luis and Haro, Josep Maria},
	month = feb,
	year = {2016},
	pages = {205--214},
}

@article{foguet2015multimorbidity,
	title = {Multimorbidity {Patterns} in {Elderly} {Primary} {Health} {Care} {Patients} in a {South} {Mediterranean} {European} {Region}: {A} {Cluster} {Analysis}},
	volume = {10},
	issn = {1932-6203},
	shorttitle = {Multimorbidity {Patterns} in {Elderly} {Primary} {Health} {Care} {Patients} in a {South} {Mediterranean} {European} {Region}},
	url = {https://journals.plos.org/plosone/article?id=10.1371/journal.pone.0141155},
	doi = {10.1371/journal.pone.0141155},
	language = {en},
	number = {11},
	urldate = {2026-09-23},
	journal = {PLOS ONE},
	publisher = {Public Library of Science},
	author = {Foguet-Boreu, Quintí and Violán, Concepción and Rodriguez-Blanco, Teresa and Roso-Llorach, Albert and Pons-Vigués, Mariona and Pujol-Ribera, Enriqueta and Gil, Yolima Cossio and Valderas, Jose M.},
	month = nov,
	year = {2015},
	pages = {e0141155},
}

@article{diaz2015multimorbidity,
	title = {Multimorbidity and {Its} {Patterns} according to {Immigrant} {Origin}. {A} {Nationwide} {Register}-{Based} {Study} in {Norway}},
	volume = {10},
	issn = {1932-6203},
	url = {https://journals.plos.org/plosone/article?id=10.1371/journal.pone.0145233},
	doi = {10.1371/journal.pone.0145233},
	language = {en},
	number = {12},
	urldate = {2026-09-23},
	journal = {PLOS ONE},
	publisher = {Public Library of Science},
	author = {Diaz, Esperanza and Poblador-Pou, Beatriz and Gimeno-Feliu, Luis-Andrés and Calderón-Larrañaga, Amaia and Kumar, Bernadette N. and Prados-Torres, Alexandra},
	month = dec,
	year = {2015},
	pages = {e0145233},
}

@book{buttorff2017multiple,
	title = {Multiple {Chronic} {Conditions} in the {United} {States}},
	isbn = {978-0-8330-9737-8},
	url = {http://www.rand.org/pubs/tools/TL221.html},
	doi = {10.7249/TL221},
	language = {en},
	urldate = {2026-09-23},
	publisher = {RAND Corporation},
	author = {Buttorff, Christine and Ruder, Teague and Bauman, Melissa},
	year = {2017},
}

@article{busija2019replicable,
	title = {Do replicable profiles of multimorbidity exist? {Systematic} review and synthesis},
	volume = {34},
	issn = {1573-7284},
	shorttitle = {Do replicable profiles of multimorbidity exist?},
	url = {https://doi.org/10.1007/s10654-019-00568-5},
	doi = {10.1007/s10654-019-00568-5},
	language = {en},
	number = {11},
	urldate = {2026-09-23},
	journal = {European Journal of Epidemiology},
	author = {Busija, Ljoudmila and Lim, Karen and Szoeke, Cassandra and Sanders, Kerrie M. and McCabe, Marita P.},
	month = nov,
	year = {2019},
	pages = {1025--1053},
}

@article{borg2017choice,
	title = {The {Choice} of {Initial} {Configurations} in {Multidimensional} {Scaling}: {Local} {Minima}, {Fit}, and {Interpretability}},
	volume = {46},
	issn = {1026-597X},
	shorttitle = {The {Choice} of {Initial} {Configurations} in {Multidimensional} {Scaling}},
	url = {https://ajs.or.at/index.php/ajs/article/view/vol46-2-2},
	doi = {10.17713/ajs.v46i2.561},
	language = {en},
	number = {2},
	urldate = {2026-09-23},
	journal = {Austrian Journal of Statistics},
	author = {Borg, Ingwer and Mair, Patrick},
	month = jan,
	year = {2017},
	pages = {19--32},
}

@incollection{balasubramanian2024interventions,
	series = {{INFORMS} {TutORials} in {Operations} {Research}},
	title = {Interventions for {Patients} with {Complex} {Medical} and {Social} {Needs}},
	isbn = {979-8-9882856-2-5},
	url = {https://pubsonline.informs.org/doi/abs/10.1287/educ.2024.0279},
	doi = {10.1287/educ.2024.0279},
	urldate = {2026-09-23},
	booktitle = {Smarter {Decisions} for a {Better} {World}},
	publisher = {INFORMS},
	author = {Balasubramanian, Hari and Prakash, Sindhoora and Jafari, Ali and Mohan, Arjun and Gopalappa, Chaitra},
	year = {2024},
	pages = {358--391},
}

@article{aoki2018multimorbidity,
	title = {Multimorbidity patterns in relation to polypharmacy and dosage frequency: a nationwide, cross-sectional study in a {Japanese} population},
	volume = {8},
	copyright = {2018 The Author(s)},
	issn = {2045-2322},
	shorttitle = {Multimorbidity patterns in relation to polypharmacy and dosage frequency},
	url = {https://www.nature.com/articles/s41598-018-21917-6},
	doi = {10.1038/s41598-018-21917-6},
	language = {en},
	number = {1},
	urldate = {2026-09-23},
	journal = {Scientific Reports},
	publisher = {Nature Publishing Group},
	author = {Aoki, Takuya and Yamamoto, Yosuke and Ikenoue, Tatsuyoshi and Onishi, Yoshihiro and Fukuhara, Shunichi},
	month = feb,
	year = {2018},
	pages = {3806},
}

@article{andrews2002deconstructing,
	title = {Deconstructing current comorbidity: data from the {Australian} {National} {Survey} of {Mental} {Health} and {Well}-{Being}},
	volume = {181},
	issn = {0007-1250, 1472-1465},
	shorttitle = {Deconstructing current comorbidity},
	url = {https://www.cambridge.org/core/journals/the-british-journal-of-psychiatry/article/deconstructing-current-comorbidity-data-from-the-australian-national-survey-of-mental-health-and-wellbeing/4CF2A5A1D28D9D5CA5E8E018A1815E4F},
	doi = {10.1192/bjp.181.4.306},
	language = {en},
	number = {4},
	urldate = {2026-09-23},
	journal = {The British Journal of Psychiatry},
	author = {Andrews, Gavin and Slade, Tim and Issakidis, Cathy},
	month = oct,
	year = {2002},
	pages = {306--314},
}

@inproceedings{agarwal1994fast,
title = {Fast Algorithms for Mining Association Rules},
author = {Agrawal, Rakesh and Srikant, Ramakrishnan},
booktitle = {Proceedings of the 20th International Conference on Very Large Data Bases ({VLDB})},
pages = {487--499},
year = {1994},
address = {Santiago, Chile},
}

@article{bjornevik_longitudinal_2022,
	address = {New York, N.Y.},
	title = {Longitudinal analysis reveals high prevalence of {Epstein}-{Barr} virus associated with multiple sclerosis},
	volume = {375},
	issn = {1095-9203},
	doi = {10.1126/science.abj8222},
	language = {eng},
	number = {6578},
	journal = {Science},
	author = {Bjornevik, Kjetil and Cortese, Marianna and Healy, Brian C. and Kuhle, Jens and Mina, Michael J. and Leng, Yumei and Elledge, Stephen J. and Niebuhr, David W. and Scher, Ann I. and Munger, Kassandra L. and Ascherio, Alberto},
	month = jan,
	year = {2022},
	pages = {296--301},
}

@dataset{ahrq_meps,
	author = {{Agency for Healthcare Research and Quality}},
	title = {{Medical Expenditure Panel Survey} household component: Medical conditions files and full-year consolidated data files, 2010--2019},
	url = {https://meps.ahrq.gov/mepsweb/data_stats/download_data_files.jsp},
}
\clearpage
\section{Appendix}
\subsection{Summary of the literature}

\setlength{\tabcolsep}{3pt}
\renewcommand{\arraystretch}{1.15}
\begin{longtable}{|>{\raggedright\arraybackslash}p{1.9cm}|>{\raggedright\arraybackslash}p{2.0cm}|>{\raggedright\arraybackslash}p{2.3cm}|>{\raggedright\arraybackslash}p{2.9cm}|>{\raggedright\arraybackslash}p{5.6cm}|}
\caption{Summary of key articles in the multimorbidity profiling literature.}
\label{tab:multimorbidity_main}\\
\hline
\textbf{Ref.} & \textbf{Country} & \textbf{Sample Size} & \textbf{Methodology} & \textbf{Key Contribution} \\
\hline
\endfirsthead
\multicolumn{5}{l}{\small\textit{Table \thetable\ -- continued from previous page}}\\
\hline
\textbf{Ref.} & \textbf{Country} & \textbf{Sample Size} & \textbf{Methodology} & \textbf{Key Contribution} \\
\hline
\endhead
\hline
\multicolumn{5}{r}{\small\textit{continued on next page}}\\
\endfoot
\hline
\endlastfoot

\cite{goldstein2009comorbidity} & USA & 3{,}595 homeless veterans & Hierarchical cluster analysis (Jaccard coefficient, Lance--Williams method) & Identified four psychiatric and general medical comorbidity clusters, noting a strong age gradient across clusters (ethnic differences were confounded by age) and substantial overlap between mental and physical illnesses. \\
\hline
\cite{poblador2014similar} & Spain and The Netherlands & Netherlands: 79{,}291; Spain: 275{,}682 & Exploratory Factor Analysis (EFA) & Compared multimorbidity patterns across two European regions, validated three consistent patterns (cardiometabolic, mechanical, psychiatric--substance abuse), and highlighted the value of electronic medical records for profiling multimorbidity. \\
\hline
\cite{diaz2015multimorbidity} & Norway & 3.7 million (Norwegians + immigrants) & Exploratory Factor Analysis (EFA), stratified by gender, age, and immigrant origin & Found substantially lower multimorbidity in immigrants than in Norwegian-born individuals (unadjusted OR 0.38--0.67 by origin), while the disease patterns themselves were broadly similar across groups. \\
\hline
\cite{foguet2015multimorbidity} & Spain & 322{,}328 elderly patients & Hierarchical cluster analysis (Ward's method, stratified by sex and age) & Found consistent circulatory--metabolic clusters and uncovered gender-specific musculoskeletal and digestive groupings. \\
\hline
\cite{garin2016global} & Spain, Finland, Poland, China, Ghana, India, Mexico, Russia, South Africa & 41{,}909 adults (50+) & Exploratory Factor Analysis (EFA) & Identified multimorbidity patterns across diverse countries, highlighting consistent cardio-respiratory and mental--articular clusters. \\
\hline
\cite{lenzi2016burden} & Italy & 3.76 million adults (Emilia-Romagna) & EFA (oblimin rotation) on administrative data & Reported high multimorbidity prevalence (15.3\% overall, 39.9\% among individuals aged \(\geq 65\) years), identified five disease patterns, and underscored the influence of citizenship, age, and gender within a universal healthcare system. \\
\hline
\cite{aoki2018multimorbidity} & Japan & 3{,}256 adults & EFA (multidimensional item response theory, promax rotation) & Identified five multimorbidity patterns and showed that the malignant/digestive/urologic and cardiovascular/renal/metabolic patterns had the strongest associations with excessive polypharmacy. \\
\hline
\cite{soley2021identifying} & UK & 826{,}166 patients (5.24 million person-years) & Longitudinal panel analysis, MCA-based multimorbidity clusters & Identified high-demand multimorbidity clusters by consultation type and ethnicity, suggesting that targeted care could reduce future primary-care strain. \\
\hline
\cite{robertson2022identifying} & Scotland & 41{,}545 patients (11{,}389 multimorbid) & Cluster analysis (Gower distance, Partitioning Around Medoids), t-SNE visualization & Demonstrated that cluster analysis can validly identify condition groupings in hospitalized patients, offering a first step toward more efficient, patient-centered care for individuals with unmet needs. \\
\hline
\cite{vaghfi2025identifying} & USA & 21.2 million de-identified patient records (Cerner Health Facts), analyzed as two random samples of 10.6 million patients each & Optimization-based clustering (mixed-integer linear programming and Bron--Kerbosch algorithm) & Developed a framework to identify lethal disease cliques in EHR data, confirming both known and novel high-mortality comorbidity patterns. \\
\hline
\end{longtable}
\normalsize

\subsection{Data preparation: imputing condition categories}

Our imputation works as follows. We consider different age groups in ten-year increments (0 to 10, 11 to 20,…, until 81+), and further subdivide individuals in each age group by gender (male and female). For each group we count all the CCCs that happened from 2010 to 2015; we show the counts for three ICD-10 codes in Table~\ref{tab:ccc_impute} for a specific age and sex group. Following that, we will go through our actual data from 2016 to 2019, so we can create a row for each individual with all their conditions in a new data set and then identify their ICD-10 codes to set the CCC column related to that ICD-10 code to 1. However, doing so might require additional steps:
\begin{enumerate}
    \item We first check whether that three-digit ICD-10 code for an individual in the MEPS 2016-19 dataset is associated with a unique CCC or multiple CCCs. As mentioned earlier, the majority of the ICD-10 3 digit codes have a unique match to the broader disease category represented by the CCC. 
    
    \item If the ICD-10 code of the patient is associated with several CCCs, then we identify that person’s age and gender and we refer to our CCC counts for that age and gender group for all CCC options we have for that ICD-10, based on the codebook. These counts are shown in Table~\ref{tab:ccc_impute}.
    
    \item We then normalize the counts for all possible CCC options for that ICD-10, as shown in the final column of Table~\ref{tab:ccc_impute}
    
    \item We sample a CCC for the ICD-10 code based on the normalized numbers that we have, i.e. the probability that the ICD-10 corresponds to a CCC in the six years before our study period (2010--2015).
    
    \item If the counts of the CCCs from 2010-2015 that map to an ICD-10 code are non-existent, we randomly choose a CCC based on the codebook (i.e. each CCC is equally likely.)
\end{enumerate}

\begin{table}[H]
\centering
\caption{Imputing a CCC from a three-digit ICD-10 code. Counts are the number
of times each candidate CCC was recorded for males aged 21--30 in MEPS
2010--2015; probabilities are those counts normalized within each ICD-10 code.
Rounded to three decimals for display; the analysis uses full precision.}
\label{tab:ccc_impute}
\setlength{\tabcolsep}{5pt}
\renewcommand{\arraystretch}{1.1}
\begin{tabular}{|l|>{\raggedright\arraybackslash}p{6.6cm}|r|r|}
\hline
\textbf{ICD-10} & \textbf{Candidate CCC} & \textbf{Count} & \textbf{Prob.} \\
\hline
\multirow{4}{*}{A20}
& Skin and subcutaneous infections (197) & 126 & 0.474 \\
& Bacterial infection, unspecified site (3) & 72 & 0.271 \\
& Pneumonia (122) & 65 & 0.244 \\
& Septicemia (2) & 3 & 0.011 \\
\hline
\multirow{12}{*}{V10}
& Cancer, other and unspecified primary (41) & 11 & 0.297 \\
& Non-Hodgkin’s lymphoma (38) & 6 & 0.162 \\
& Other non-epithelial skin cancer (23) & 5 & 0.135 \\
& Melanoma of skin (22) & 4 & 0.108 \\
& Tuberculosis (1) & 2 & 0.054 \\
& Cancer of other GI organs (18) & 2 & 0.054 \\
& Cancer of prostate (29) & 2 & 0.054 \\
& Cancer of colon (14) & 1 & 0.027 \\
& Cancer of bronchus/lung (19) & 1 & 0.027 \\
& Cancer of testis (30) & 1 & 0.027 \\
& Cancer of other male genital organs (31) & 1 & 0.027 \\
& Thyroid cancer (36) & 1 & 0.027 \\
\hline
\multirow{4}{*}{C45}
& Cancer, other and unspecified primary (41) & 11 & 0.647 \\
& Tuberculosis (1) & 2 & 0.118 \\
& Cancer of other GI organs (18) & 2 & 0.118 \\
& Malignant neoplasm, site unspecified (43) & 2 & 0.118 \\
\hline
\end{tabular}
\end{table}

\subsection{Stable clusters from hierarchical clustering}

For each integer $k$ in the range $k = 3, 4, \dots, 50$, we used the \texttt{maxclust} criterion to obtain a partition of the diseases into up to $k$ clusters from the Ward dendrogram. Each partition can be viewed as a collection of clusters (sets of diseases).

However, in hierarchical clustering, increasing $k$ by one does not always change the underlying partition: it is possible for the clustering at $k$ and $k+1$ to be identical. To avoid double-counting such cases, we compared each new partition to the previous one. If the entire set of clusters (as sets of disease codes) was unchanged relative to the previous $k$, we did not update any counts. If we are able to form $k+1$ clusters, it means that one of the original clusters has split while the others have remained intact. This split is an indicator of the relative strength and stability of the clusters that persist across different cut levels. Only when the partition truly changed did we:
\begin{enumerate}
    \item record each cluster as a set of disease codes
    \item increment a global counter tracking how many times that exact cluster appeared across the grid of $k$ values.
\end{enumerate}

This procedure yields a ``cluster stability database'' in which each row corresponds to a unique disease cluster, together with:
\begin{itemize}
    \item its length (number of diseases in the cluster), and
    \item the number of distinct partitions, across $k = 3, \dots, 50$, in which it occurs.
\end{itemize}

\subsection{Final clusters for men aged 40--60}

\begin{figure}[H]
  \centering
  \includegraphics{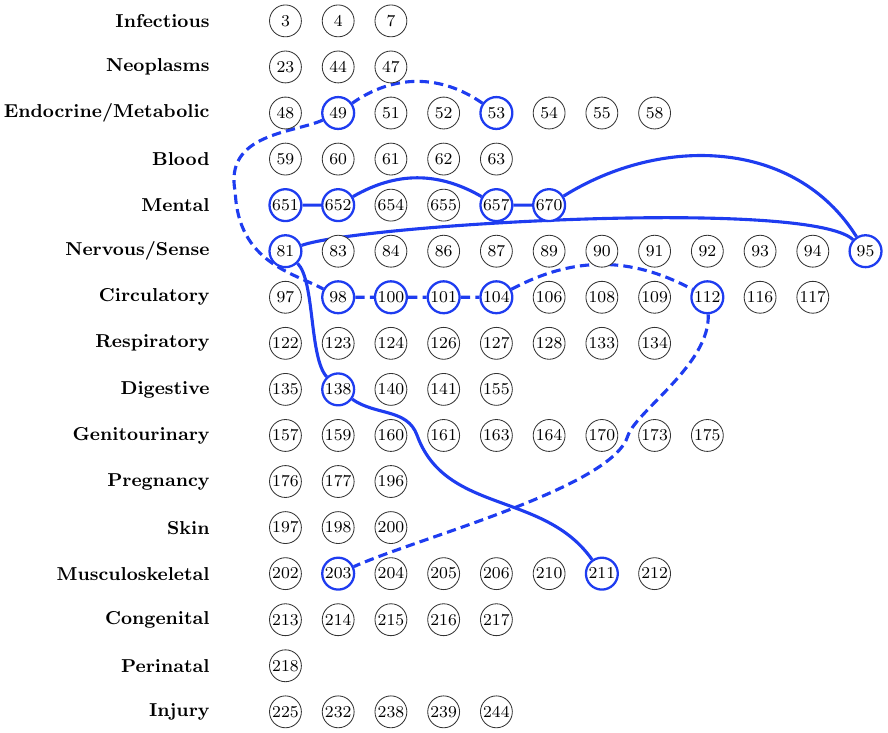} 
  \caption{Final disease cluster examples—Men, 40–60. Each line joins the conditions of one cluster (thick circles); Table~\ref{bodysystems} gives the disease name for each code. Drawn with TikZ code written with the help of Anthropic’s Claude Opus 5.5 and checked by the authors. }
  \label{fig:cluster-men-4060}
\end{figure}

The first cluster for men aged $40-60$ spans four different body systems. The solid line in Figure  \ref{fig:cluster-men-4060} connects all the 8 diseases in this cluster. In this visualization, the diseases are organized in rows, with each row corresponding to the 16 different body systems. Each row consists of numbers in circles which refer to the diseases within that body system. Thus the cluster represented by the solid line connects four diseases from Mental Health (Anxiety Disorders (651), Attention-Deficit/Conduct/Disruptive Behaviors (652), Mood Disorders (657) and Miscellaneous Mental Health Disorders (670)); two diseases from Nervous System and Sense Organs (Other Hereditary and Degenerative Nervous System Conditions (81), Other Nervous System Disorders (95)), Digestive (Esophageal Disorders (138)), and Musculoskeletal (Connective Tissue disorders (211)) systems. Similarly, the second cluster, represented by a dashed line in Figure~\ref{fig:cluster-men-4060}, connects diseases from the Endocrine/Metabolic, Circulatory, and Musculoskeletal systems. This cluster contains common conditions such as diabetes without complications (49), high cholesterol or lipid metabolism disorders (53), and high blood pressure or hypertension (98) and well known ``downstream'' diseases that are far more serious, such as heart attack or acute myocardial infarction (100), plaque formation in the heart also known as coronary atherosclerosis (101), other heart disease (104), and mini-stroke or transient ischemic attack (112). The cluster also includes one musculoskeletal disease, Osteoarthritis (203). Both are eight-disease clusters; the solid-line cluster was returned in 25 of the 100 iterations and the dashed cluster in 8, the two highest frequencies among eight-disease clusters in this stratum.

\subsection{Clusters by age group, and the body-system mapping}

Tables~\ref{men_byage} and~\ref{women_byage} give three or four of the most frequently occurring clusters of three or more diseases in each age
group, and Table~\ref{bodysystems} gives the mapping from condition categories to body systems used to
compute Diversity.

\begin{table}[H]
\centering
\caption{Male disease clusters by age group with disease names, diversity values, and frequencies. Numbers in parentheses are CCC codes; Table~\ref{bodysystems} lists them by body system.}
\label{men_byage}
\begin{tabular}{|c|p{8.5cm}|c|c|}
\hline
\textbf{Age Group} & \textbf{Cluster (Disease Name + Code)} & \textbf{Diversity} & \textbf{Frequency} \\
\hline

\multirow{3}{*}{0–20} 
& Back problem (205), Ot conn tiss (211), Anxiety disorders (651), Attention-deficit/conduct/disruptive behavior disorders (652), Disorders usually diagnosed in infancy or childhood (655), Mood disorders (657) & 2 & 26 \\
\cline{2-4}
& Oth bact inf (3), Viral infect (7), Ot up rsp in (126), COPD (127), Oth low resp (133) & 2 & 22 \\
\cline{2-4}
& Viral infect (7), Ot up rsp in (126), COPD (127), Oth low resp (133) & 2 & 20 \\
\hline

\multirow{3}{*}{20–40} 
& Headache/mig (84), Blindness (89), Influenza (123), Asthma (128) & 2 & 24 \\
\cline{2-4}
& DiabMel no c (49), Hyperlipidem (53), Hypertension (98), Ot uppr resp (134), Esophgeal dx (138) & 4 & 21 \\
\cline{2-4}
& DiabMel no c (49), Hyperlipidem (53), Hypertension (98), Esophgeal dx (138) & 3 & 18 \\
\hline

\multirow{3}{*}{40–60} 
& Thyroid dsor (48), Ot endo dsor (51), Gout/ot crys (54) & 1 & 54 \\
\cline{2-4}
& Ot uppr resp (134), Joint injury (225), Other injury (244) & 2 & 32 \\
\cline{2-4}
& DiabMel no c (49), Hyperlipidem (53), Acute MI (100), Coronary atherosclerosis and other heart disease (101), Oth heart dx (104), TIA (112), Osteoarthritis (203) & 3 & 27 \\
\hline

\multirow{3}{*}{60+} 
& Viral infect (7), Ot up rsp in (126), Sprain (232), Superfic inj (239), Other injury (244) & 3 & 40 \\
\cline{2-4}
& Ot dx stomch (141), Complic proc (238), Anxiety disorders (651), Miscellaneous mental health disorders (670) & 3 & 27 \\
\cline{2-4}
& Teeth dx (136), Ot dx stomch (141), Complic proc (238) & 2 & 22 \\
\hline

\end{tabular}
\end{table}

\begin{table}[H]
\centering
\caption{Female disease clusters by age group with disease names, diversity values, and frequencies. Numbers in parentheses are CCC codes; Table~\ref{bodysystems} lists them by body system.}
\label{women_byage}
\begin{tabular}{|c|p{8.5cm}|c|c|}
\hline
\textbf{Age Group} & \textbf{Cluster (Disease Name + Code)} & \textbf{Diversity} & \textbf{Frequency} \\
\hline
\multirow{3}{*}{0–20} 
& Ot uppr resp (134), Menstrual dx (171), Back problem (205) & 3 & 21 \\
\cline{2-4}
& Mycoses (4), Nutrit defic (52), Headache/mig (84), Ot dx stomch (141), Contraceptiv (176), Anxiety disorders (651), Mood disorders (657) & 6 & 21 \\
\cline{2-4}
& Ot dx stomch (141), Contraceptiv (176), Mood disorders (657) & 3 & 20 \\
\hline
\multirow{3}{*}{20–40} 
& Spont abortn (177), Other pregnancy and delivery including normal (196), Liveborn (218) & 2 & 37 \\
\cline{2-4}
& DiabMel no c (49), Hyperlipidem (53), Oth nerv dx (95), Hypertension (98), Esophgeal dx (138), Osteoarthritis (203), Ot conn tiss (211) & 5 & 27 \\
\cline{2-4}
& DiabMel no c (49), Hyperlipidem (53), Ot nutrit dx (58), Hypertension (98), Ot joint dx (204) & 3 & 23 \\
\hline
\multirow{3}{*}{40–60} 
& Blindness (89), Urinary tract infections (159), Ot femal gen (175) & 2 & 30 \\
\cline{2-4}
& Acute MI (100), Coronary atherosclerosis and other heart disease (101), Oth heart dx (104), Art embolism (116) & 1 & 22 \\
\cline{2-4}
& Fluid/elc dx (55), Intest infct (135), Esophgeal dx (138), Biliary dx (149), Other GI dx (155) & 2 & 20 \\
\hline
\multirow{4}{*}{60+} 
& Gastritis (140), Urinary tract infections (159), Prolapse of female genital organs (170) & 2 & 28 \\
\cline{2-4}
& Dizziness (93), Ot circul dx (117), Sprain (232) & 3 & 25 \\
\cline{2-4}
& Ot bnign neo (47), Retinal dx (87), Other ear dx (94) & 2 & 25 \\
\cline{2-4}
& Non-epith ca (23), Nutrit defic (52), Headache/mig (84), Ot infl skin (198), Ot bone dx (212) & 5 & 22 \\
\hline
\end{tabular}
\end{table}

\FloatBarrier
\begin{longtable}{|c|>{\raggedright\arraybackslash}p{11.4cm}|}
\caption{Mapping from CCC condition categories to the sixteen Level-1 body systems used to compute Diversity,
with the AHRQ Clinical Classifications Software (CCS) short names used throughout the paper. All sixteen body systems are listed so that the table can
be read as a general reference; none of the clusters shown in this paper contains a condition from the Blood (4) or
Congenital (14) groups.}
\label{bodysystems}\\
\hline
\textbf{Group (Code)} & \textbf{Diseases (Name + Code)} \\
\hline
\endfirsthead
\hline
\textbf{Group (Code)} & \textbf{Diseases (Name + Code)} \\
\hline
\endhead
Infectious (1) & Mycoses (4), Oth bact inf (3), Viral infect (7) \\
\hline
Neoplasms (2) & Neoplsm unsp (44), Cancer of prostate (29), Non-epith ca (23), Ot bnign neo (47) \\
\hline
Endocrine/Metabolic (3) & DiabMel no c (49), Diabetes (with complications) (50), Fluid/elc dx (55), Gout/ot crys (54), Hyperlipidem (53), Nutrit defic (52), Ot endo dsor (51), Ot nutrit dx (58), Thyroid dsor (48) \\
\hline
Blood (4) & Anemia (59), Acute posthemorrhagic anemia (60), Sickle cell anemia (61), Coag/hemr dx (62), Wht blood dx (63) \\
\hline
Mental Health (5) & Anxiety disorders (651), Attention-deficit/conduct/disruptive behavior disorders (652), Developmental disorders (654), Disorders usually diagnosed in infancy or childhood (655), Miscellaneous mental health disorders (670), Mood disorders (657) \\
\hline
Neuro/Sensory (6) & Blindness (89), Glaucoma (88), Cataract (86), Dizziness (93), Epilepsy/cnv (83), Eye infectn (90), Headache/mig (84), Ot hered CNS (81), Oth nerv dx (95), Other ear dx (94), Other eye dx (91), Otitis media (92), Retinal dx (87) \\
\hline
Circulatory (7) & Acute CVD (109), Acute MI (100), Art embolism (116), Cardiac dysrhythmias (106), Carditis (97), Coronary atherosclerosis and other heart disease (101), Hypertension (98), Ot circul dx (117), Oth heart dx (104), TIA (112), chf;nonhp (108) \\
\hline
Respiratory (8) & Asthma (128), COPD (127), Influenza (123), Ot up rsp in (126), Ot uppr resp (134), Oth low resp (133), Pneumonia (122), Tonsillitis (124) \\
\hline
Digestive (9) & Esophgeal dx (138), Teeth dx (136), Biliary dx (149), Gastritis (140), Intest infct (135), Ot dx stomch (141), Other GI dx (155) \\
\hline
Genitourinary (10) & Acute and unspecified renal failure (157), Other male genital disorders (166), Menstrual dx (171), BPH (164), Menopausl dx (173), Ot dx kidney (161), Ot femal gen (175), Other GU dx (163), Prolapse of female genital organs (170), Urin stone (160), Urinary tract infections (159) \\
\hline
Pregnancy (11) & Contraceptiv (176), Other pregnancy and delivery including normal (196), Spont abortn (177) \\
\hline
Skin (12) & Ot infl skin (198), Oth skin dx (200), Skin infectn (197) \\
\hline
Musculoskeletal (13) & Back problem (205), Osteoarthritis (203), Osteoporosis (206), Ot bone dx (212), Ot conn tiss (211), Ot joint dx (204), Rheum arth (202), SLE (210) \\
\hline
Congenital (14) & Cardiac anom (213), GI cong anom (214), GU cong anom (215), Nerv cong an (216), Other congenital anomalies (217) \\
\hline
Perinatal (15) & Liveborn (218) \\
\hline
Injury (16) & Complic proc (238), Fracture of arm (229), Joint injury (225), Other injury (244), Sprain (232), Superfic inj (239) \\
\hline
\end{longtable}

\end{document}